\documentclass[final,5p,times,twocolumn, %authoryear
]{elsarticle}

\usepackage{listings}
\usepackage{array}
\usepackage{amssymb}
\usepackage{lipsum}
\usepackage{amsmath}
\usepackage[hidelinks]{hyperref}
\usepackage{listings}
\usepackage{paralist}
\setcitestyle{numbers}
\usepackage{makecell}
\usepackage{rotating}
\usepackage{amssymb}
\usepackage{booktabs} % For pretty tables
\usepackage{subfigure}
\usepackage{lipsum}
\usepackage{xcolor}
\usepackage{hyperref}
\usepackage{graphicx}
\usepackage{tcolorbox}
\usepackage{balance}
\usepackage[T1]{fontenc}

\newcommand{\toolname}{CyberRAG}

\usepackage{float}

\usepackage{amsmath}

\begin{document}

\begin{frontmatter}
\journal{Future Generation Computer Systems}


\title{CyberRAG: An Agentic RAG  cyber attack classification and reporting tool} 


\author[1,2]{Francesco Blefari\corref{cor1}}
\cortext[cor1]{Corresponding author}
\ead{francesco.blefari@unical.it}
\author[1]{Cristian Cosentino}
\ead{cristian.cosentino@unical.it}
\author[1]{Francesco Aurelio Pironti}
\ead{francesco.pironti@unical.it}
\author[1]{Angelo Furfaro}
\ead{angelo.furfaro@unical.it}
\author[1]{Fabrizio Marozzo}
\ead{fabrizio.marozzo@unical.it}
\affiliation[1]{organization={University of Calabria}, 
%Department and Organization
            addressline={Via Pietro Bucci}, 
            city={Rende},
            postcode={87036}, 
            state={Italy},
            country={}}

\affiliation[2]{organization={IMT School for Advanced Studies},
%Department and Organization
            addressline={Piazza San Francesco}, 
            city={Lucca},
            postcode={55100}, 
            state={Italy},
            country={}}

\begin{abstract}
Intrusion Detection and Prevention Systems (IDS/IPS) in large enterprises can generate hundreds of thousands of alerts per hour, overwhelming analysts with logs requiring rapidly evolving expertise. Conventional machine-learning detectors reduce alert volume but still yield many false positives, while standard Retrieval-Augmented Generation (RAG) pipelines often retrieve irrelevant context and fail to justify predictions. We present \toolname{}, a modular agent-based RAG framework that delivers real-time classification, explanation, and structured reporting for cyber-attacks. A central LLM agent orchestrates: (i) fine-tuned classifiers specialized by attack family; (ii) tool adapters for enrichment and alerting; and (iii) an iterative retrieval-and-reason loop that queries a domain-specific knowledge base until evidence is relevant and self-consistent. Unlike traditional RAG, \toolname{} adopts an agentic design that enables dynamic control flow and adaptive reasoning. This architecture autonomously refines threat labels and natural-language justifications, reducing false positives and enhancing interpretability. It is also extensible: new attack types can be supported by adding classifiers without retraining the core agent. \toolname{} was evaluated on SQL Injection, XSS, and SSTI, achieving over 94\% accuracy per class and a final classification accuracy of 94.92\% through semantic orchestration. Generated explanations reached 0.94 in BERTScore and 4.9/5 in GPT-4-based expert evaluation, with robustness preserved against adversarial and unseen payloads. These results show that agentic, specialist-oriented RAG can combine high detection accuracy with trustworthy, SOC-ready prose, offering a flexible path toward partially automated cyber-defense workflows.
\end{abstract}


\end{frontmatter}

%\tableofcontents



\section{Introduction}
\label{introduction}
%decidere se è un tool o un sistema.

The cybersecurity landscape has advanced considerably, moving from manual expert-driven processes to increasingly automated and intelligent systems~\cite{mohamed2025}. However, the interpretation and response to cyber threats remains largely semi-automated and dependent on human expertise, particularly in large-scale enterprise environments. Intrusion Detection and Prevention Systems (IDS/IPS) continue to be foundational to network defense~\cite{LIAO2013}, but they generate massive volumes of alerts, often hundreds of thousands per hour, many of which require expert validation. Although machine learning-based detectors can reduce this burden, they often suffer from high false positive rates and limited interpretability~\cite{blefari2024}. 

Moreover, the output produced by current IDS/IPS is typically presented as raw log strings: highly informative, yet difficult to read and interpret. Logs are static artifacts that lack interactivity and do not provide the possibility for clarification or contextualization. As a result, when an alert is triggered, highly skilled analysts must manually investigate the underlying event, potentially diverting the attention of security teams from truly critical threats.

A promising direction to address the aforementioned challenges lies in leveraging LLMs to support the analysis of logs and alerts, enhancing both the readability of results and the analyst’s ability to make informed decisions. In recent years, large language models (LLMs) have gained traction in cybersecurity due to their ability to interpret threat data and support analysts through natural language reasoning~\cite{llm_cybersecurity_sota2024}. A key advancement in this area is RAG~\cite{lewis2020retrieval}, which enriches model input with relevant context retrieved from external data sources. This combination improves the quality and grounding of responses~\cite{yamin2024applications}. Still, most RAG implementations retrieve context only once before generating output, lacking the ability to refine queries, reason iteratively, or dynamically adapt to complex situations.

To push beyond these limitations, autonomous AI agents have emerged as a promising paradigm. These agents are designed to autonomously perform tasks such as continuous monitoring, anomaly detection, and threat mitigation, increasingly becoming a key component of modern security operations~\cite{deng2025ai,behera2025artificial}. When powered by LLMs, such agents can not only interpret security data but also coordinate tools, issue actions, and generate reports. However, many LLM-based agents still operate as black boxes, making decisions without exposing their rationale.

To overcome the limitations of conventional RAG pipelines and the opacity of LLM-based agents in cybersecurity, we introduce \toolname{}, a modular and extensible agent-based RAG framework for real-time cyber-attack classification, explanation, and reporting. \toolname{} is designed to address two critical requirements: 
\begin{inparaenum}[\itshape(i)\upshape]
    \item task specialization, through a set of fine-tuned LLM classifiers, each targeting a specific attack category (e.g., DDoS, ransomware, SQL injection); and 
    \item context-aware reasoning, enabled by a multi-phase RAG component that iteratively retrieves and refines relevant information from a domain-specific knowledge base. 
\end{inparaenum}
At the center of the framework is a large language model acting as an autonomous agent that orchestrates the classification pipeline, invokes specialized tools as needed, and generates structured, interpretable reports. Unlike standard RAG systems that retrieve context in a single pass, \toolname{} allows the agent to reason over retrieved evidence and re-query the knowledge base to refine its understanding.

Designed to integrate seamlessly with existing IDS infrastructures, \toolname{} activates upon receiving a flagged alert and autonomously processes the associated network traffic. It identifies the likely attack category, retrieves supporting contextual information, and generates a comprehensive report in natural language that describes the threat and suggests mitigation steps. An integrated LLM-powered assistant enables interactive querying of the report, providing analysts with deeper insights or recommended remediation actions. The system is easily customizable: organizations can expand its internal knowledge base using internal documents, architectural diagrams, or policy files, without the need to fine-tune the agent or classifiers.

%These results demonstrate that an agent-based, specialist-driven RAG framework can deliver both high detection performance and interpretable, Security Operations Center (SOC) ready outputs, offering a practical step toward semi-autonomous cybersecurity operations.

The remainder of this paper is organized as follows:
Section~\ref{sec:background} introduces the topic by presenting foundational cybersecurity concepts, such as web-based attacks, and important LLMs notions including AI agents and Agentic RAG.
Section~\ref{sec:relatedwork} discusses the related work.
Our methodology is detailed in Section~\ref{sec:methodology}, followed by a discussion presenting the Knowledge Bases and the datasets in Section~\ref{sec:kbds}. The results are shown in Section~\ref{sec:results} and discussed in Section~\ref{sec:discussion}.
Finally, conclusions are drawn in Section~\ref{sec:conclusion}.

% \begin{figure}
%     \centering
%     \includegraphics[width=\linewidth]{scenario.pdf}
%     \caption{Internal flow of CyberRAG: the user query triggers specialized modules combining classifiers and attack description tools to generate responses.}
%     \label{fig:scenario}
% \end{figure}

\section{Background}
\label{sec:background}
The growing interconnection of web services has made the detection and mitigation of sophisticated cyber-attacks increasingly critical. Organizations today face threats ranging from credential theft to large-scale offensives, such as Distributed Denial of Service (DDoS)~\cite{FURFARO2020} attacks and Advanced Persistent Threats (APTs). Among the most critical vulnerabilities, web applications represent a prime target for malicious actors who exploit browser flaws, hijack user sessions, or inject harmful code.

% The tool we propose is a custom-built solution designed to assist security operators in identifying and resolving vulnerabilities that commonly arise during normal system operations. It leverages well-known web-based attack vectors to simulate real-world threats and incorporates a collection of established concepts such as Large Language Models (LLMs), Intrusion Detection Systems, Agents and Agentic RAG which serve as foundational pillars for our approach.

\toolname{} leverages well-known web-based attack vectors to simulate real-world threats and incorporates a collection of established concepts such as LLMs, IDS, AI Agents and Agentic RAG technology which serve as foundational pillars for our approach.
These technologies are central to the objectives we aim to achieve and directly influence the design and implementation of our tool. In the following, we provide an overview of the key technologies and methodologies referenced throughout this work.

\subsection{Intrusion Detection and Prevention Systems}
An \textit{Intrusion Detection System} (IDS) is a crucial component in cybersecurity for identifying ongoing attacks. It functions by monitoring network traffic from the infrastructure it aims to protect, and comparing that traffic against a database of known threats~\cite{debar2009}. The two primary detection approaches are signature-based and anomaly-based methods.
The \textit{signature-based} approach identifies intrusions by matching network data against predefined patterns, such as specific keywords in HTTP requests or known malicious byte sequences. In contrast, the \textit{anomaly-based} approach leverages machine learning techniques to detect previously unseen threats. These systems learn a model of normal network behavior and raise alerts when deviations from this baseline are observed, potentially indicating malicious activity.
When a threat is detected, the IDS generates an alert to notify security operators, who are then responsible for deciding the most appropriate response strategy. Hybrid approaches, aiming at exploiting the advantages of both approaches are known in literature~\cite{Angiulli2018}.
An \textit{Intrusion Prevention System} (IPS) extends the functionality of an IDS by not only detecting threats but also actively preventing them. Positioned logically, and sometimes physically, between protected hosts and external networks, the IPS intercepts malicious traffic and takes immediate action, such as blocking the connection or isolating affected systems. It uses the same detection techniques as IDS but includes automated response capabilities.
A common drawback of IPS solutions is their potential to become a single point of failure in the network, which can impact availability if the system itself is compromised or misconfigured.

\subsection{Cyber-attacks}
Web-based attacks are among the most dangerous forms of cyber-attacks, as they allow unprivileged users, such as casual visitors to a website, to gain an initial foothold within a corporate network. Once inside, attackers can exploit this access to perform lateral movements across systems, gradually escalating their privileges.
By chaining multiple stages of lateral movement and privilege escalation, an attacker can eventually gain full control over the entire system. The ultimate objective of these attacks is either to seize control of the target machine, typically through code injection techniques like SSTI, or to exfiltrate sensitive data, as seen in SQL injection and Cross-Site Scripting (XSS) attacks.
These representative attack vectors have been selected to evaluate the capabilities of our tool and are discussed in detail in the remainder of this section.

\subsubsection{Cross-Site Scripting (XSS)}
Cross-Site Scripting (XSS) attacks are a type of security vulnerability that leverages JavaScript code injection~\cite{gupta2017, HYDARA2015}, allowing an external unauthenticated entity to execute the injected code in the victim's web browser. The primary goal of this attack is to steal sensitive information, such as passwords, credit card details, or cookies.  

XSS attacks exploit the trust that a web browser places in the content received from a server. Since browsers expect to receive only text, an attacker must find a way to send malicious input to the server, which is then parsed and transmitted back to the browser as part of a web page. When the browser renders this response, it encounters injected HTML tags containing JavaScript code, which it mistakenly executes as legitimate commands. For instance, an attacker might insert malicious code inside \texttt{<script></script>} tags to execute arbitrary JavaScript.  

There are three main types of XSS attacks:  

\begin{itemize}
    \item \textit{Reflected XSS attacks}, also known as non-persistent XSS, typically involve a specially crafted URL that an attacker tricks a user into visiting. The malicious script is included in the URL and reflected by a vulnerable web page, such as an error page or a search results page, before being executed in the user's browser.  
    \item \textit{Stored XSS attacks}, also known as persistent XSS, occur when a web application stores user-provided input without properly sanitizing it. In these attacks, an attacker injects malicious JavaScript into a website's database (e.g., via a comment field or a user profile). When another user visits the affected page, the script is retrieved from storage and executed in their browser.  
    \item \textit{DOM-based XSS} occurs when an attacker manipulates the Document Object Model (DOM) of a web page using client-side scripts. Unlike other XSS attacks, the malicious payload does not travel through the server; instead, it is executed by modifying the DOM in the user's browser, exploiting vulnerabilities in JavaScript execution.  
\end{itemize}

\subsubsection{Server-Side Template Injection (SSTI)}  
Server-Side Template Injection (SSTI) is a vulnerability that arises when an attacker is able to inject malicious payloads into a server-side template, exploiting the template engine’s capabilities to execute unauthorized code on the server~\cite{Zhao2023}.  

Many modern web applications utilize template engines to dynamically generate HTML by embedding user inputs into predefined templates. If the application fails to properly sanitize or validate these inputs, the template engine may interpret and execute them as code. This can lead to serious security consequences, including unauthorized data access, exfiltration of sensitive information, or even full remote code execution (RCE) on the server.  

Commonly used template engines known to be susceptible to SSTI include:  
%\begin{inparaitem}  
    %\item 
    Jinja2 (Python),  
    %\item 
    Twig (PHP),  
    %\item 
    Mako (Python),  
    %\item 
    FreeMarker (Java),  
    %\item 
    Velocity (Java).
%\end{inparaitem}  

Each of these engines has unique syntax and capabilities, but the underlying vulnerability remains the same: improper handling of untrusted user input. Attackers often start by testing for SSTI through the use of basic expressions. For instance, in the case of Jinja2, an attacker might input the string \texttt{\{\{7*7\}\}}. If the rendered output from the server contains \texttt{49}, it confirms that the expression was evaluated, thus revealing the presence of an exploitable SSTI vulnerability.

\subsubsection{SQL Injection}
A SQL injection is an attack that occurs when an attacker is able to insert and execute malicious code within a vulnerable SQL query, as discussed in~\cite{Singh2016}. These attacks are commonly used to bypass login mechanisms or to exfiltrate sensitive information from databases.

The primary cause of SQL injection vulnerabilities lies in the use of unsanitized user inputs directly within SQL queries, combined with the lack of secure programming practices such as prepared statements or parameterized queries.

Classical forms of SQL injection include \textit{Piggy-Backed Queries}, where the attacker manipulates input fields by inserting malicious SQL code, often using characters like the semicolon ``\verb+;+'' to terminate a legitimate query and append a malicious one. Another typical case is the exploitation of \textit{Stored Procedures}, where attackers create or manipulate database procedures through injected SQL statements.

More advanced forms of SQL injection include:

\begin{itemize}
    \item \textit{Blind SQL Injection}, used when the application does not return visible error messages but allows inference through changes in behavior. This can be further classified into:
    \begin{enumerate}
        \item \textit{Boolean-based (content-based)}: the attacker sends boolean conditions in queries and observes the difference in content or structure of the application’s response to deduce information.
        \item \textit{Time-based}: the attacker leverages commands such as \texttt{SLEEP()} or \texttt{WAITFOR DELAY} to induce time delays, inferring data from the duration of the response.
\end{enumerate}     
    \item \textit{Union-based SQL Injection}, in which the attacker uses the SQL \texttt{UNION} operator to combine the result of the original query with another malicious query, effectively retrieving data from other tables.

    \item \textit{Out-of-Band SQL Injection}, used when traditional techniques fail. This method relies on external systems like DNS or HTTP requests to exfiltrate data. It is particularly effective when the attacker cannot observe the application's response directly but can trigger events on a remote server.
\end{itemize}

\subsection{Large Language Models}
Large Language Models have become one of the most significant advances in Natural Language Processing (NLP), with growing applications in areas such as cybersecurity, medicine, and law~\cite{brown2020language,devlin2019bert,radford2019language,vaswani2017attention,liu2019roberta}. These models are typically based on the Transformer architecture~\cite{vaswani2017attention}, which makes it possible to capture long-range dependencies through self-attention mechanisms. Their training usually follows a two-phase paradigm: a large-scale pre-training stage on general-purpose corpora, followed by task-specific fine-tuning on curated datasets. This combination enables LLMs to learn general language representations while remaining adaptable to specialized domains.  

From an architectural perspective, LLMs are commonly distinguished into two broad families: encoder-based models and decoder-based models. While they share the same Transformer foundation, their objectives and downstream applications differ substantially.  
Taken together, the two families highlight complementary strengths: encoder models excel at precise and context-aware classification, while decoder models provide detailed and interpretable explanations. 

\subsubsection{Encoder-based Models}
Encoder-based models, exemplified by BERT~\cite{devlin2019bert} and RoBERTa~\cite{liu2019roberta}, are primarily designed for discriminative tasks that require a bidirectional understanding of text. By processing input in both directions, they capture fine-grained semantic relations, which makes them particularly effective for classification, semantic matching, and anomaly detection. In our system, such models are employed as specialized classifiers, each focusing on a distinct type of cyber-attack (e.g., SQL Injection, XSS, SSTI). This specialization allows the classifiers to capture the unique linguistic patterns and syntactic features associated with each attack vector.  

\subsubsection{Decoder-based Models}
Decoder-based models, such as GPT~\cite{brown2020language} and the LLaMA family~\cite{touvron2024llama}, follow an autoregressive objective, predicting the next token in a sequence. This makes them naturally suited for generative tasks such as text completion, explanation synthesis, and narrative construction. 
Within our architecture, decoder-based models are integrated into the RAG pipeline, where they generate context-aware descriptions of attacks by combining classifier outputs with relevant external knowledge.
%In this way, they support the production of natural-language justifications \evb{and reports ready for SOC operators} that complement the predictions of encoder-based models.  

%By combining them, our approach aims to maximize both accuracy and interpretability, thereby addressing the dual requirement of detection performance and analyst trust in SOC environments~\cite{hendrycks2020pretrained,ribeiro2020beyond}.

While proprietary frontier models such as GPT-4 demonstrate remarkable generative capabilities, they cannot be fine-tuned on domain-specific corpora due to the lack of access to their weights. In highly specialized domains such as cybersecurity, this limits their applicability to zero-shot or few-shot prompting, which generally yields lower and less stable results~\cite{cosentino2024exploiting,cantini2024multi}. For this reason, we focus on open-weight encoder models that can be adapted and replicated, while keeping the framework compatible with future open-weight frontier models.

\subsection{AI Agents}
An Agent is a computer system situated in an environment that can autonomously act in its context to reach its delegated objectives~\cite{wooldridge:2009}. 

\begin{figure*}[ht]
    \centering
    \includegraphics[width=.65\linewidth]{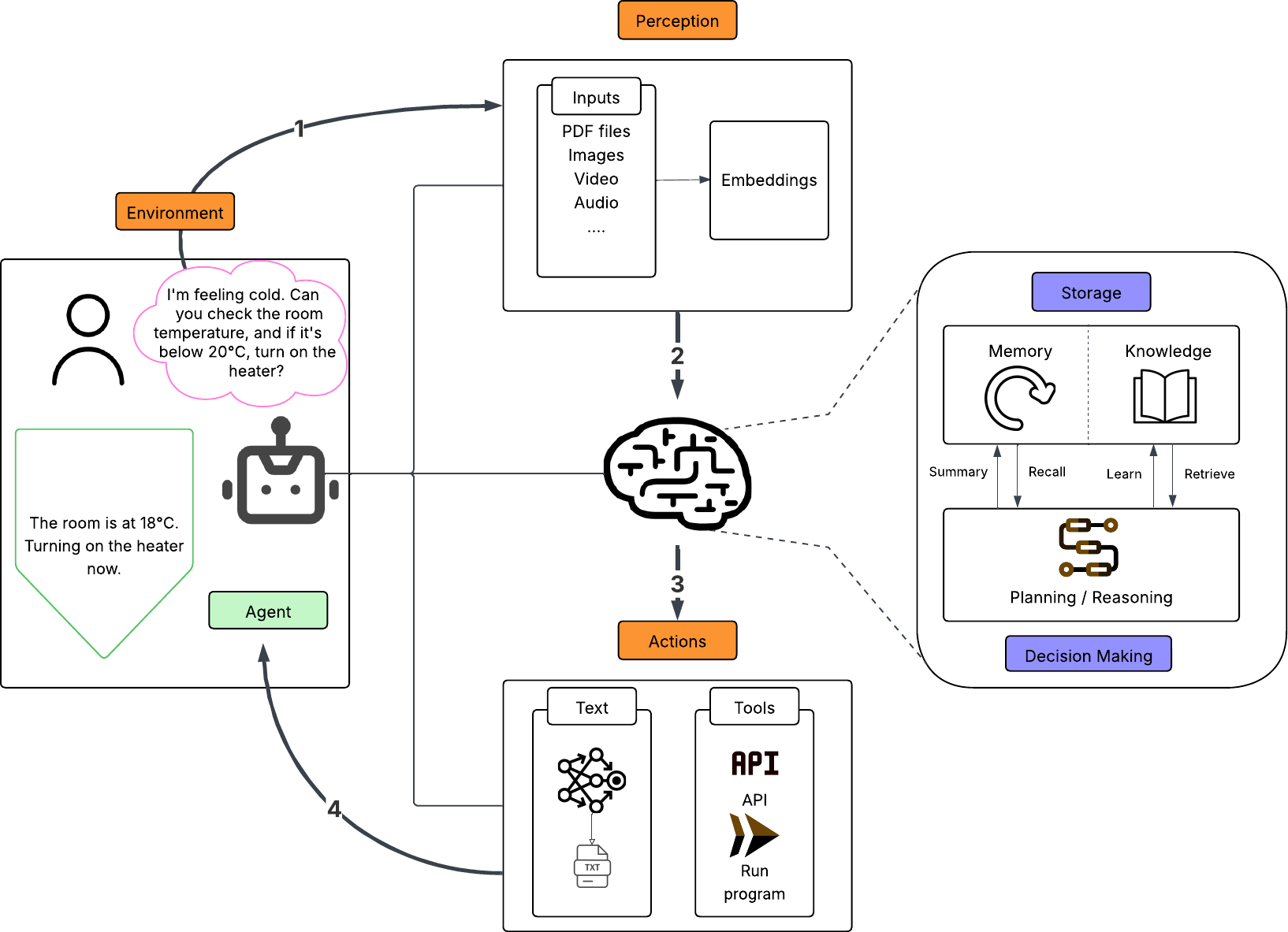}
    \caption{Structure of a multimodal AI Agent: integrates perception, memory, and planning to interact with the environment and perform intelligent actions.}
    \label{fig:AI_agent_schema}
\end{figure*}

The term \textit{autonomy} refers to the capability and requisites necessary to determine the appropriate course of action to achieve a specified objective. An intelligent agent (AI Agent) is characterized by its ability to perceive its environment, respond to changes within it, initiate actions, and engage with other systems, which may include other agents or human users. One of the core principles foundational to AI agents is the concept of \textit{memory}. Effective memory management improves an agent's capability in maintaining contextual awareness. It enables the agent to draw on previous experiences effectively, thereby facilitating the development of incrementally informed decision-making abilities as time progresses.
These concepts are depicted in Figure~\ref{fig:AI_agent_schema}.
As noted in~\cite{LLM_Agents_survey}, the emergence of LLMs represents another moment of progress in the realization of AI agents. Substantial advances in this direction have culminated in the emergence of LLM agents. In particular, these agents use LLM as reasoning and planning cores to decide the control flow of an application, while maintaining the characteristics of traditional AI agents. LLM agents extend the capabilities of LLMs by facilitating the use of external tools to address specialized tasks, including mathematical computation and code execution. Ultimately, the model evaluates the adequacy of its output and determines whether additional processing is necessary.

\subsection{Agentic RAG}
Although general-purpose LLMs and their associated agents offer remarkable versatility, they frequently lack the \textit{in-depth domain-specific} expertise essential for addressing intricate and specialized problems. One potential solution to this limitation involves retraining or fine-tuning the model; however, undertaking such procedures can be both financially demanding and resource consuming. An alternative strategy to circumvent these challenges emerges from the RAG paradigm. This strategy enhances LLMs by connecting them to external domain-specific knowledge sources, such as document collections, through a retriever component. Functioning like a search engine, the retriever extracts relevant information from this external repository and integrates it directly into the LLM's context. This provides the model with the necessary data to effectively answer complex requests.

In RAG systems, external data is first loaded and segmented into appropriately sized chunks. These chunks are then converted into vector representations (embeddings) and stored in specialized data structures for future retrieval. The typical workflow of a RAG system follows these steps:
\begin{inparaenum}[(i)]
    \item the user submits a query to the system;
    \item the retriever converts the query into a vector representation and searches for the most relevant stored embeddings, retrieving the corresponding chunks;
    \item 
    the original query is enriched with the retrieved chunks and passed to the LLM;
    \item the LLM generates a context-aware response, which is then presented to the user.
\end{inparaenum}

An agent system that utilizes the RAG paradigm is commonly referred to as an Agentic RAG. Several state-of-the-art frameworks, including LangChain~\cite{LangChain}, LlamaIndex~\cite{LlamaIndex}, and Langdroid~\cite{Langdroid}, offer user-friendly interfaces for building customized Agentic RAG solutions.

\subsection{Attack classification and detection techniques}
\label{sec:relatedwork}

%The growing interconnection of web services has made the detection and mitigation of sophisticated cyber-attacks increasingly critical. Organizations today face threats ranging from credential theft to large-scale offensives, such as Distributed Denial of Service (DDoS)~\cite{FURFARO2020} attacks and Advanced Persistent Threats (APTs). Among the most critical vulnerabilities, web applications represent a prime target for malicious actors who exploit browser flaws, hijack user sessions, or inject harmful code.

%Traditional defense mechanisms, based on static rules or known signatures, prove inadequate in the face of rapidly evolving intrusion techniques and the increasing frequency of zero-day attacks. This necessitates the development of more intelligent and adaptive solutions, capable not only of detecting known threats, but also of continuously learning from and adapting to novel attack patterns.

%In this context, language models such as BERT~\cite{devlin2019bert} offer a powerful opportunity: their ability to understand complex linguistic contexts allows for flexible and generalizable modeling of attack descriptions. Our approach leverages this capability to enable the dynamic integration of more types of attack without requiring a complete system redesign.
In Table~\ref{tab:comparison_techniques} some key difference between classical and AI-powered techniques are proposed.

\begin{table*}[t]
\footnotesize
\centering
\caption{Comparison between Classical and AI-powered Techniques for Cyber-Attack Detection}
\label{tab:comparison_techniques}
\begin{tabular}{p{3.5cm} p{5.5cm} p{5.5cm}}
\hline
\textbf{Aspect} & \textbf{Classical Techniques} & \textbf{AI-powered Techniques (LLMs)} \\
\hline
\textbf{Detection Approach} & Signature-based, rule-driven, taxonomy-defined & Context-aware, data-driven, language understanding \\

\textbf{Adaptability to Novel Attacks} & Low - limited to known patterns or signatures & High - capable of generalization and zero-day inference \\

\textbf{Application Context} & Static systems, IDS/IPS, network-layer security & Dynamic classification, forensic analysis, attack generation and simulation \\

\textbf{Attack Scope} & Focused on DoS, DDoS, passive/active, and layered attacks & Includes phishing, malware, ransomware, social engineering, nation-state threats \\

\textbf{Granularity of Analysis} & Protocol- and layer-specific, domain-expert encoded & Textual and semantic inference, able to correlate across diverse data types \\

\textbf{Key Challenges} & Limited scalability, rigid taxonomies, high false positives & Prompt injection risks, adversarial robustness, interpretability \\
\hline
\textbf{Example Tools / Frameworks} & Firewall, IPS/IDS & AutoAttacker, SecBench \\
\hline
\end{tabular}
\end{table*}

\begin{table*}[t]
\footnotesize
\centering
\caption{Summary of Related Works in Cybersecurity Detection and Classification}
\label{tab:related_works_summary}
\begin{tabular}{p{2.8cm}p{4.5cm}p{4.5cm}p{4cm}}
\hline
\textbf{Work} & \textbf{Focus Area} & \textbf{Main Contribution} & \textbf{Remarks} \\
\hline
\textbf{AVOIDIT~\cite{simmons2009avoidit}} & Taxonomy of cyber attacks & Proposes a structured classification of attacks by vector, objective, and impact & Foundational for holistic threat modeling \\

\textbf{Howard et al.~\cite{howard2007modern}} & Intrusion Detection Systems & Explores modern IDS/IPS strategies & Emphasizes signature-based defenses \\

% \textbf{Shimeall et al.~\cite{shimeall2002cyberterrorism}} & Cyberterrorism & Discusses the strategic and socio-political impact of cyber attacks & Highlights need for broader defense frameworks \\

\textbf{AutoAttacker~\cite{autoattacker2024}} & LLM-driven cyber offense & Simulates AI-powered attack chains using GPT-based modules in enterprise settings & Demonstrates offensive use of LLMs \\

\textbf{RAG for Cyber Defense~\cite{rag_cyberattack2024}} & Retrieval-Augmented Generation & Integrates threat intelligence into LLM-based classification via retrieval mechanisms & Improves context relevance and detection accuracy \\

\textbf{ChatAPT~\cite{chatapt2024}} & Nation-state threat attribution & Uses LLMs with threat intelligence and knowledge graphs for campaign attribution & Supports advanced threat actor profiling \\
\hline
\end{tabular}
\end{table*}

\subsubsection{Classical techniques}
% descrizione delle tecniche di rilevamento classiche presenti in letteratura

In the context of cyber attack classification and detection systems, ``classical'' techniques generally focus on the definition of taxonomies, attack patterns, and well-established defensive methods. A traditional aspect involves distinguishing between passive attacks (e.g., eavesdropping and traffic analysis) and active attacks (e.g., injecting malicious packets or tampering with data). %Additionally, attacks are often classified into the macro-categories of \emph{cyber crime}, \emph{cyber espionage}, \emph{cyber terrorism}, and \emph{cyber war}, highlighting how objectives and consequences can vary significantly.

An important example of classification is the AVOIDIT taxonomy, which groups attacks according to vectors, objectives, and impacts, underscoring the need for a holistic perspective on security~\cite{simmons2009avoidit}. Within this framework, threats such as Denial of Service (DoS), port scanning, unauthorized data exfiltration, and, more broadly, data integrity violations are analyzed. In parallel, the literature has also focused on distributed attacks (DDoS), where multiple compromised nodes act simultaneously to deny service or cause damage with high impact~\cite{Anshuman2022}.

%In the context of mobile ad hoc networks (MANETs) and wireless sensor networks (WSNs), “classical” classification strategies emphasize the detection of threats such as black hole and wormhole attacks, as well as Byzantine variations that compromise nodes internally~\cite{awerbuch2004mitigating}. 
%Awerbuch et al.~\cite{awerbuch2004mitigating} offer one of the foundational references to defend against Byzantine attacks in wireless environments, introducing measures aimed at detecting and isolating malicious nodes. Other studies have emphasized the distinction between single-layer and multi-layer attacks, in which adversaries exploit vulnerabilities at various levels of the communication stack~\cite{cheung2003modeling}.

On the prevention and response front, many works focus on the integration of cryptographic tools, authentication systems, and key-management protocols to mitigate the risk of device compromise~\cite{cashell2004economic}.
%,cleveland2010cyber}. 
This approach often assumes that the primary defensive perimeter is defined by cryptographic robustness and the proper configuration of network services, prioritizing the use of IDS and IPS~\cite{howard2007modern,jovicic2006common}. 
%However, as documented in reports on targeted attacks and more complex scenarios such as cyber espionage or cyber terrorism, there is a need for a broader perspective that accounts for socio-technical and strategic aspects~\cite{
%shimeall2002cyberterrorism,vijayan2010targeted}.

\subsubsection{AI-powered classification and defense techniques}
% descrizione delle tecniche di rilevamento degli attacchi web che utilizzano intelligenza artificiale (ml, dl, gen-ai).

Recent research underscores the growing impact of AI-based tools, particularly Large Language Models , in the domain of cyber-attack orchestration and defense. Early studies demonstrated that LLMs such as ChatGPT could generate attack scripts with success rates ranging from 16\% up to 50\% when combined with basic cybersecurity skills, thereby lowering the barrier to entry for malicious actors~\cite{autoattacker2024,generative_ai_cybersecurity2024}. Offensive platforms like WormGPT and FraudGPT were explicitly developed to harness LLM capabilities for malicious objectives such as phishing, ransomware development, and malware generation~\cite{llm_cybersecurity_review2024}. %These specialized LLMs highlight a pivotal shift in the cyber-threat landscape, where automation of sophisticated exploits is no longer confined to highly skilled adversaries.

At the same time, the cybersecurity community has begun to leverage LLMs in a defensive capacity, exploiting their ability to interpret contextual nuances in textual data~\cite{CHEN2024104016,generative_ai_cybersecurity2024}. Advanced frameworks have been proposed to harness LLMs in automated attack classification, attribution, and system hardening. One illustrative example is the \textit{AutoAttacker} system, which uses LLM-guided modules to orchestrate and evaluate attacks on a simulated organizational network~\cite{autoattacker2024}. 
%Meanwhile, works such as \textit{CyberSecEval 2} and \textit{SecBench} emphasize benchmarking and comparative evaluations of LLM-based methods to measure their robustness against real-world offensive scenarios~\cite{cyberseceval2024,secbench2024}.

%Despite these advances, multiple studies argue that conventional benchmarks fail to capture the full spectrum of real-world risks posed by LLMs in cybersecurity settings~\cite{llm_cyber_risk2025}.
%For instance, emerging methods of \emph{prompt injection} have exposed how easily an LLM’s responses can be manipulated to generate misleading or harmful content~\cite{prompt_injection_defense2024}. In response, new lines of research have proposed refining LLMs with domain-specific corpora and advanced retrieval-augmentation techniques (often denoted as RAG: Retrieval-Augmented Generation) to provide more accurate and context-aware threat assessments~\cite{rag_cyberattack2024,ctikg2024}.

Studies focusing on threats at the \emph{nation-state} level also point to the increasing importance of LLM-driven analysis in attributing sophisticated attacks~\cite{chatapt2024}.
%Here, combining LLM capabilities with knowledge graphs and advanced threat intelligence fosters deeper insights into malware campaigns and social engineering tactics~\cite{llm_cybersecurity_sota2024}. 
Furthermore, reviews of cyber defense LLM applications highlight how these models can enhance intrusion detection, automate forensic analysis, and generate real-time alerts, although persistent challenges remain in interpretability, adversarial robustness, and regulatory compliance~\cite{llm_cybersecurity_review2024,generative_ai_cybersecurity2024}.

A summary of the works discussed above is provided in Table~\ref{tab:related_works_summary}.
%Within this evolving landscape, \toolname{} draws on the strengths of LLMs to assist in classifying and reporting cyber-attacks while mitigating adversarial manipulations.

\section{Methodology}
\label{sec:methodology}

\begin{figure*}[ht]
    \centering
    \includegraphics[width=\linewidth]{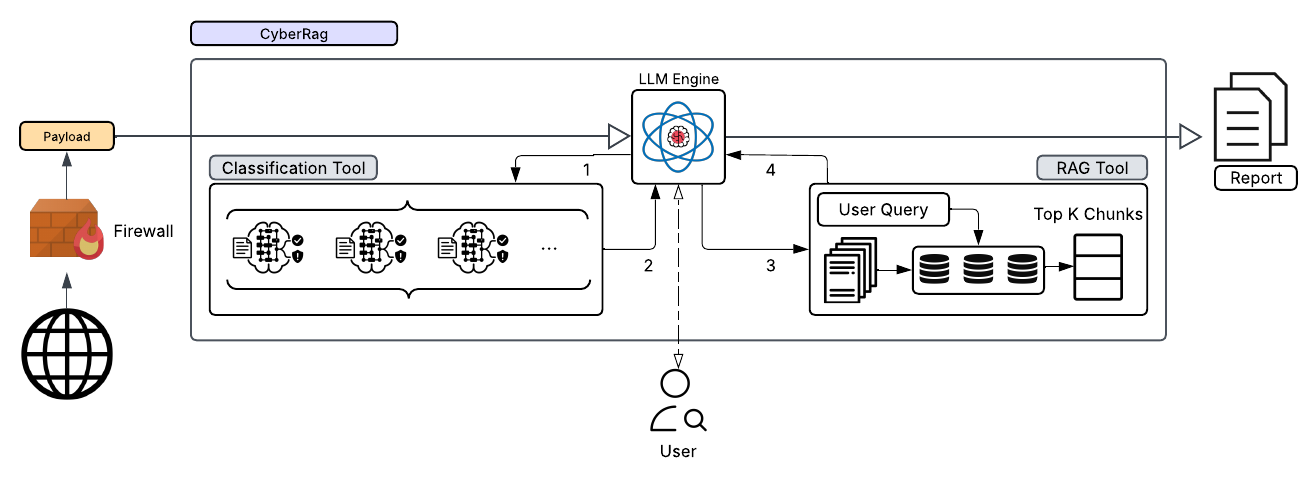}
    \caption{\toolname{} system architecture: the user interacts with a chatbot connected to the webserver, while the IDS detects attacks from the Internet.}
    \label{fig:enter-label}
\end{figure*}

The growing sophistication of cyber-attacks necessitates detection systems that are not only accurate but also comprehensible and adaptable. To address these demands, \toolname{} leverages Agentic RAG technology. \toolname{} has a modular architecture whose main component is a general purpose LLM, here referred to as  the \textit{core LLM Engine}. This LLM  is in charge of handling the overall \toolname{} operation. It directly performs some actions and exploits two other components (tools) for the execution of some specific tasks.  The core LLM, by acting as the central intelligent agent that orchestrates the entire system, dynamically manages the two tools and invokes the services offered by other component modules. Each component (tool or module) is responsible for a distinct analytical task, such as classification, contextualization, or user interaction, enabling the Agentic RAG system to dynamically orchestrate reasoning steps in response to uncertainty or ambiguity. This modular design, directed by the core LLM Engine, enhances robustness, scalability, and interpretability.

In contrast to traditional RAG approaches that rely on a single retrieval step, the Agentic RAG framework supports multiple iterative retrieval passes. This allows the system, under the guidance of the Core LLM Engine, to autonomously reassess and refine its initial decisions. For example, if the initial classification is ambiguous or incorrect, \toolname{} can reclassify the input and re-query the appropriate knowledge base, thereby improving both the coherence and accuracy of the final explanation.
%More specifically, \toolname{} leverages two core components -- namely, the \textit{Classification} tool and the \textit{RAG} tool -- to detect malicious payloads. Moreover, by harnessing the intrinsic capabilities of LLMs, it generates a detailed report describing the detected attacks and also enables users to query the information for targeted insights.
The \toolname{}'s architecture, depicted in Figure~\ref{fig:enter-label}, highlights its fundamental components. %These include the \textit{Classification} tool, responsible for detecting and categorizing incoming payloads; the \textit{RAG} tool, which enriches the classification with external knowledge. \toolname{} also leverages LLM capabilities to provide description and deep dive explanation of the analyzed payloads. More in detail, \toolname{} is also featured by the \textit{Attack Description Report Generator}, which synthesizes the final narrative explanation; and the \textit{User Chat } interface, which enables human-interpretable interaction and exploration.

The \textit{Classification} tool is in charge of payload classification by employing a set of specific LLM-based classifiers, each fine-tuned on a given attack class (e.g. SQL Injection, XSS, SSTI). Although this work focuses on three classes, \toolname{} is designed for extensibility, allowing the addition of new attack categories with minimal effort.

\toolname{} uses the \textit{RAG tool} to generate a high-quality informative attack description by  employing the relevant knowledge associated with the detected attack class. 
%The information collected is then processed, leveraging the generative capabilities of its core LLM, to produce detailed and contextualized explanations of the vulnerability.
%As a result, \toolname{} not only classifies attacks, but also provides a comprehensive, human-readable narrative report that significantly enhances understanding.
%
The gathered information is then processed by leveraging the generative capabilities of the core LLM Engine, which transforms technical insights about the detected vulnerability into detailed and contextualized explanations. As a result, \toolname{} not only classifies attacks, but also provides a comprehensive, human-readable narrative report that significantly enhances understanding using the \textit{Attack Description Report Generator} module.

When the core LLM exhibits low confidence in its classification or encounters ambiguity, users can initiate a dialogue with the system (similar to interacting with a chatbot). This enables the system to engage in focused questioning, requesting clarification or additional information before proceeding. This interactive mechanism improves both the accuracy and the interpretability of the system, particularly in complex or edge-case scenarios.

In the following sections, each of these tools and core modules are described in detail. We explain their internal mechanisms, how they interact, and how they contribute to the overall pipeline of cyber-attack interpretation and reporting.

\subsection{Classification Tool}
Once the core LLM engine receives the payload for analysis, it leverages the classification tool to determine the appropriate cyber-attack family to which the payload belongs.

The classification of cyber-attack payloads is managed through a modular ensemble of models based on the BERT family of transformers. Each model in this ensemble is individually fine-tuned to specialize in recognizing a specific type of cyber-attack, enabling targeted detection across a wide range of threat categories.

The fine-tuning process involves adapting a pre-trained language model, initially trained on a large, general-purpose corpus, to a more specific task: identifying particular attack patterns. Through this task-specific training, each model learns to extract and interpret features that are highly relevant to its assigned attack class, enhancing both the precision and robustness of classification, even when payloads exhibit subtle structural variations or deliberate obfuscation.

Within this architecture, each specialized model functions as an independent semantic classifier. Given an input payload, a model produces a classification label indicating the predicted attack type, a confidence score reflecting the certainty of the prediction, and an explanatory component.

The \textit{Classification} tool, to determine the most plausible attack class evaluates the confidence scores produced by each specialized classifier. Rather than applying abstract reconciliation strategies such as majority voting or threshold-based filtering, the LLM-driven decision process directly leverages the highest confidence value as the primary indicator. The class with the highest score is selected as the most reliable prediction and is used to condition the construction of the knowledge-grounded prompt for the retrieval module.

The output of all classifiers is aggregated into a unified structured comparison table. This table enables parallel evaluation of the same payload from multiple semantic perspectives, thus exploiting the complementarity of the specialized models. Such a design ensures scalability and modularity, allowing the addition of other classifiers without disrupting existing components.

The structured output (see Table~\ref{tab:classification_output_clean}), is then further processed by the core LLM engine as detailed in the following.

% Poi nel corpo del documento:
\renewcommand{\arraystretch}{1.2}
\begin{table*}[htb]
\centering
\caption{Structured output from the \textit{Classification Tool} (prediction scores per class for each payload).}
\label{tab:classification_output_clean}
\fontsize{8pt}{9pt}\normalfont{
\begin{tabular}{ccccc}
\toprule
\textbf{ID} & \textbf{Payload} & \textbf{SQLi} & \textbf{SSTI} & \textbf{XSS} \\
\midrule
\textbf{PD001} &
\shortstack[l]{\texttt{1' and 3580 = ( select count ( * ) from domain.domains} \\ 
\texttt{as t1, domain.columns as t2, domain.tables as t3 ) $--$}} 
& 0.9999 & 0.3956 & 0.0673 \\

\\
\textbf{PD002} &
\shortstack[l]{\texttt{1''\}\}\{\{1016814002+3472965455\}\} \{\{'bo'\}\}} \\
\texttt{\{\#comment\#\}\{\% raw 'bo'.join('7n') \%\}} \\
\texttt{\{\{'7n'\}\}\{\{3140320242+4078937248\}}} 
& 0.3999 & 0.9997 & 0.3830 \\

\\
\textbf{PD003} &
\texttt{<time onpointermove=alert(1)>XSS</time>} 
& 0.3998 & 0.3929 & 0.9999 \\
\bottomrule
\end{tabular}
}
\end{table*}

\subsection{RAG Tool}
This component enhances the interpretability and informativeness of the system's output by starting from the structured intermediate representation, encapsulating the core features, justifications, and contextual metadata of the attack, produced by the \textit{Classification} tool, and integrating external knowledge sources through the \textit{Retrieval-Augmented Generation} (RAG) mechanism.

The core LLM engine leverages the semantic representation to automatically generate a natural language query that reflects the identified attack type, salient payload characteristics, and relevant contextual indicators. This query is submitted to a semantic search engine that indexes curated cybersecurity resources, including:
\begin{inparaenum}[\itshape(i)\upshape]
\item vulnerability databases (e.g., Common Vulnerabilities and Exposures \href{https://www.cve.org/}{CVE}, National Vulnerability Database \href{https://nvd.nist.gov/}{NVD}),
\item technical documentation (e.g., Open Worldwide Application Security Project \href{https://owasp.org/}{OWASP}, \href{https://attack.mitre.org/}{MITRE ATT\&CK}),
\item scientific literature and incident reports.
\end{inparaenum}

To retrieve relevant documents, the system performs a similarity search using dense vector representations stored in three distinct in-memory vector stores, each optimized for a specific source domain. Among the different  retrieval strategies, we adopted \textbf{Maximal Marginal Relevance (MMR)} due to its ability to balance relevance and diversity~\cite{carbonell1998use}. This ensures that the retrieved documents are not only topically relevant to the query but also non-redundant, thereby providing broader contextual coverage.

The top-ranked documents are then summarized, and a narrative contextualization of the attack payload is generated. This includes common usage patterns, associated CVEs, threat severities, and recommended mitigation strategies. The goal is to bridge the gap between low-level payload analysis and high-level cybersecurity knowledge, supporting both automated agents and human analysts.

\subsection{Attack Description and Report Generation module}
After the payload is classified by the \textit{Classification} tool and relevant information is retrieved by the \textit{RAG} tool, an \textit{Attack Description Report} is generated by leveraging the descriptive capabilities of the core LLM. Aiming to produce a comprehensive narrative of the attack, \toolname{} synthesizes information coming from both the \textit{Classification} tool %, which provides the inferred attack type and its underlying justifications, 
and the \textit{RAG} tool.%, which contributes broader contextual knowledge derived from external cybersecurity resources.

%In cases of uncertainty—where multiple classes exhibit close or competing confidence values—the system performs prompt replication: issuing the retrieval query twice with slight variations. This approach helps reduce ambiguity and maximize the informativeness of the retrieved content, leading to a richer and more complete representation of the incident.

Once the retrieval phase is complete, the core LLM engine composes a structured semantic summary of the incident. This process builds an \textit{attack representation} that captures several key aspects, including:

\begin{itemize}
\item the inferred attack type and a justification derived from the model's reasoning;
\item salient features of the input payload that contributed to the classification;
\item contextual indicators, such as script patterns, DOM elements, or specific input field characteristics that signal malicious behavior;
\item high-level knowledge retrieved from external sources, including typical usage scenarios, associated CVEs, threat severity, and recommended mitigation strategies.
\end{itemize}

This representation is then refined to generate a human-readable attack description. It also provides a solid foundation for downstream tasks such as autonomous decision-making by agent-based systems or integration into reporting and logging pipelines.

By abstracting and reasoning over the outputs of both the classification and retrieval components, this phase significantly enhances system interpretability, supports traceability, and enables explainable decision-making. At the end of this process, a report containing detailed information about the payload is produced.

\subsection{User Chat}
To complement the automated pipeline, the system integrates an interactive \textit{Chat} module that leverages the intrinsic capabilities of the core LLM engine to support natural language exploration of the results. This component allows users to ask questions about classifications, detection mechanisms, and underlying reasoning, thereby enhancing both transparency and interpretability.

The chat interface dynamically adapts its responses based on the user's input and the enriched semantic content generated by the RAG module. It supports:
\begin{inparaenum}[\itshape(i)\upshape]
\item explanations of the classification rationale;
\item elaborations on attack vectors and behavioral patterns;
\item response adaptation based on the user's level of expertise, from novice to advanced;
\item interactive guidance on how to patch the identified vulnerability.
\end{inparaenum}

\paragraph{Example: chatbot QA interaction} In the following example, the user engages with the chatbot to gain a clearer understanding of why a specific payload was classified as a Server-Side Template Injection (SSTI) attack.
\begin{tcolorbox}[colback=cyan!5!white,colframe=cyan!50!black,title=\textbf{Chatbot interaction}]
\small
\textbf{User:} Why was this classified as SSTI?

\textbf{System:} The payload \texttt{{{7*7}}} is indicative of Server-Side Template Injection (SSTI). It reflects a typical exploit pattern where unescaped user input is evaluated in a template engine, such as Jinja2. Based on semantic similarity and historical pattern recognition, the system assigned a high confidence classification. Supporting information was retrieved from OWASP references and CVE-2020-17526, confirming the vulnerability pattern.
\end{tcolorbox}
This interactive dialogue layer enhances system transparency, fostering user trust, comprehension, and overall usability—particularly in high-stakes cybersecurity contexts where interpretability of automated decisions is critical.

\paragraph{Example: SQL Injection Case} The following example illustrates the report generated by \toolname{} after performing the various analysis phases when processing the SQL Injection payload shown below.

\begin{tcolorbox}[colback=blue!3!white,colframe=blue!80!black,title=\textbf{Payload}]
\footnotesize
\ttfamily 1\%"" ) ) waitfor delay '0:0:5' and    (    (   ""\%""  =  ""\}
\end{tcolorbox}

\begin{tcolorbox}[colback=white!5!white,colframe=black!60!black,title=\textbf{Report}]

\begin{tcolorbox}[colback=yellow!5!white,colframe=yellow!60!black,title=\textbf{Analytical Summary}]
\small
\begin{enumerate}[(1)]
\item \textbf{String Encapsulation:} The presence of single quotes indicates attempts to break or inject into SQL string boundaries.
\item \textbf{Command-like Patterns:} The use of waitfor delay suggests an attempt to delay response, commonly used in blind SQL injection techniques.
\item \textbf{Syntax Anomalies:} Redundant parentheses and erratic spacing are indicative of non-standard inputs aimed at bypassing filters or altering parser logic.
\end{enumerate}
\end{tcolorbox}

\begin{tcolorbox}[colback=green!5!white,colframe=green!50!black,title=\textbf{Conclusion}]
\small
The payload presents a high-confidence match with known SQL injection signatures. This determination is supported by both syntactic irregularities and functional behaviors that mirror typical exploitation patterns. The classification is further validated through corroborating evidence drawn from authoritative threat intelligence sources.
\end{tcolorbox}

\begin{tcolorbox}[colback=purple!3!white,colframe=purple!50!black,title=\textbf{Feature Vector Summary}]
\small
\begin{itemize}
\item \textbf{SQL\_keywords\_count:} 0
\item \textbf{Dynamic\_values\_count:} High
\item \textbf{SQL\_syntax\_match:} Low
\end{itemize}
\end{tcolorbox}

\end{tcolorbox}

Although typically hidden, some LLM engines provide access to their internal reasoning processes. In this example, DeepSeek-R1:7B is used, and its reasoning is presented below.

\begin{tcolorbox}[colback=gray!5!white,colframe=black!75!white,title=\textbf{Reasoning (core LLM)}]
\small
The system identifies this payload as indicative of a SQL injection attack. The reasoning process highlights multiple suspicious elements: the use of single quotes that typically denote string delimiters in SQL, the presence of a time delay command intended to alter query execution flow, and structurally malformed syntax with excessive and misaligned parentheses. These traits commonly appear in obfuscated or malicious payloads designed to manipulate database queries.
\end{tcolorbox}

\section{Knowledge Bases and Datasets}
\label{sec:kbds}
\subsection{Knowledge Bases}

For the purposes of this work, we constructed three distinct knowledge bases, one for each type of web-based attack under investigation.
%, and an additional one for storing contextual information such as internal procedures and documentation. 
The system is designed to distinguish among three specific attack types: SSTI, SQL Injection, and Cross-Site Scripting (XSS). To achieve a deep understanding of these attack vectors, dedicated and categorized documents are required.

We sourced our documentation from PortSwigger’s \href{https://portswigger.net/web-security}{Web Security Academy}, a well-established educational platform in web security. All relevant materials were downloaded and archived as PDF files. These documents were then organized into folders corresponding to their respective attack categories: \texttt{SSTI}, \texttt{SQL Injection}, and \texttt{XSS}.

Once populated, each knowledge base was processed to enable semantic retrieval. This involved splitting the documents into smaller, manageable chunks. The size of each chunk was set to \texttt{800} characters with an overlap of \texttt{80} characters to preserve context across boundaries. Each chunk was subsequently transformed into \textit{embeddings}, using a pre-trained Sentence Transformer model, in this instance Sentence-BERT~\cite{Reimers2019SentenceBERTSE}.

These embeddings were stored in efficient vector databases, such as in-memory vector stores, using the Faiss library~\cite{douze2025faisslibrary}, which pairs each chunk with its corresponding embedding. This structure allows for rapid and contextually relevant retrieval based on semantic similarity to user queries or system needs.

\subsection{Attacks' Dataset}
Given the absence of a unified dataset tailored to our specific scenario, we opted to construct a custom dataset by aggregating and adapting existing publicly available resources. Our sources include selected Kaggle datasets~\cite{XSS, sstidataset} and the GitHub repository \href{https://github.com/swisskyrepo/PayloadsAllTheThings}{PayloadsAllTheThings}, which is widely recognized for its comprehensive collection of real-world attack payloads.

More specifically, for the \textit{SQL Injection} and \textit{Cross-Site Scripting (XSS)} categories, we utilized well-structured datasets available on Kaggle~\cite{sqlinj, XSS}. These datasets provide both positive samples (malicious payloads) and negative samples (benign inputs), enabling a supervised learning approach for classification tasks.  

For the \textit{Server-Side Template Injection (SSTI)} category, no ready-to-use dataset is available. Consequently, we generated this dataset manually using curated payloads from the PayloadAllTheThings SSTI section in conjunction with outputs from the \href{https://github.com/vladko312/SSTImap?tab=readme-ov-file}{SSTImap} tool. The positive class consists of diverse payloads known to trigger SSTI vulnerabilities across multiple template engines. In contrast, the negative class comprises a variety of benign strings such as mail, names or numbers that do not result in SSTI behavior, serving as clean input examples. This dataset can be downloaded by password at~\cite{sstidataset} an example is provided in Table~\ref{tab:ssti_table}.
\begin{table}[]
    \centering
    \scriptsize
    \begin{tabular}{|c|c|}
    \hline
    payload & label \\
    \hline
        \{\{\{ ''.\_\_class\_\_.\_\_mro\_\_[1].\_\_subclasses\_\_()[59]('/etc/passwd').read() \}\}\} & 1 \\

       (3a + 1z) + (9b * 9) = 0 & 0 \\
        \hline
         
    \end{tabular}
    \caption{Extract of SSTI dataset}
    \label{tab:ssti_table}
\end{table}
%This custom dataset  construction allows us to emulate realistic attack scenarios while maintaining balanced and well-labeled data distributions across the different attack types.

\section{Experimental Results}
\label{sec:results}
This section presents the experimental evaluation of \toolname{}. The goal is to validate: the effectiveness of employing an Agentic RAG approach, the benefits of integrating a retrieval mechanism, and the ability of modern open-weight language models to generate accurate and interpretable threat descriptions.

More in-depth,  each individual component of \toolname{} has been validated, and an assessment of the performance and reliability of the whole system has been performed. This includes the evaluation of the accuracy of the LLM-based classifiers, the effectiveness of the RAG-based explanation module, and the coherence and usability of the final outputs generated by the complete Agentic RAG pipeline.

%\begin{itemize} \item[(i)] In \ref{sec:results:agents}, we assess the classification performance of the specialized agents, each trained to detect a specific attack type (SQL Injection, SSTI, and XSS), and compare it to a baseline monolithic model. \item[(ii)] In \ref{sec:results:rag_orch}, we introduce a RAG-based orchestration mechanism that refines ambiguous classifications using domain knowledge and contextual reasoning. \item[(iii)] In \ref{sec:results:rag_eval}, we evaluate the quality of explanations generated by different RAG pipelines using a variety of automatic natural language metrics. \item[(iv)] In \ref{sec:results:ablation}, we perform an ablation study to quantify the impact of the retrieval component on the quality of generated justifications. \item[(v)] In \ref{sec:results:robustness}, we test system robustness against adversarial perturbations and out-of-distribution (OOD) inputs. \item[(vi)] In \ref{sec:results:learning_strategies}, we compare different learning strategies (zero-shot, few-shot, fine-tuned) to highlight the trade-offs in classification accuracy. \item[(vii)] Finally, in \ref{sec:results:llm_judge}, we present an external evaluation of the generated explanations using GPT-4 as an LLM-based judge, simulating expert-level assessment. \end{itemize}

\subsection{LLM-based Classifier for Attack Identification}
\label{sec:results:agents}
To improve the classification of web-based attacks, we designed a system of three independent LLMs, each dedicated to the detection of a specific attack type: SQL Injection, SSTI, and XSS. Rather than relying on a single general-purpose classifier, this architecture allows each model to specialize in learning the nuances, patterns, and syntactic/semantic features of a particular attack category.

\begin{figure}[!htb]
\centering
\subfigure[{\scriptsize SQL Injection.}\label{fig:bert_sql}]
{\includegraphics[width=0.48\textwidth]{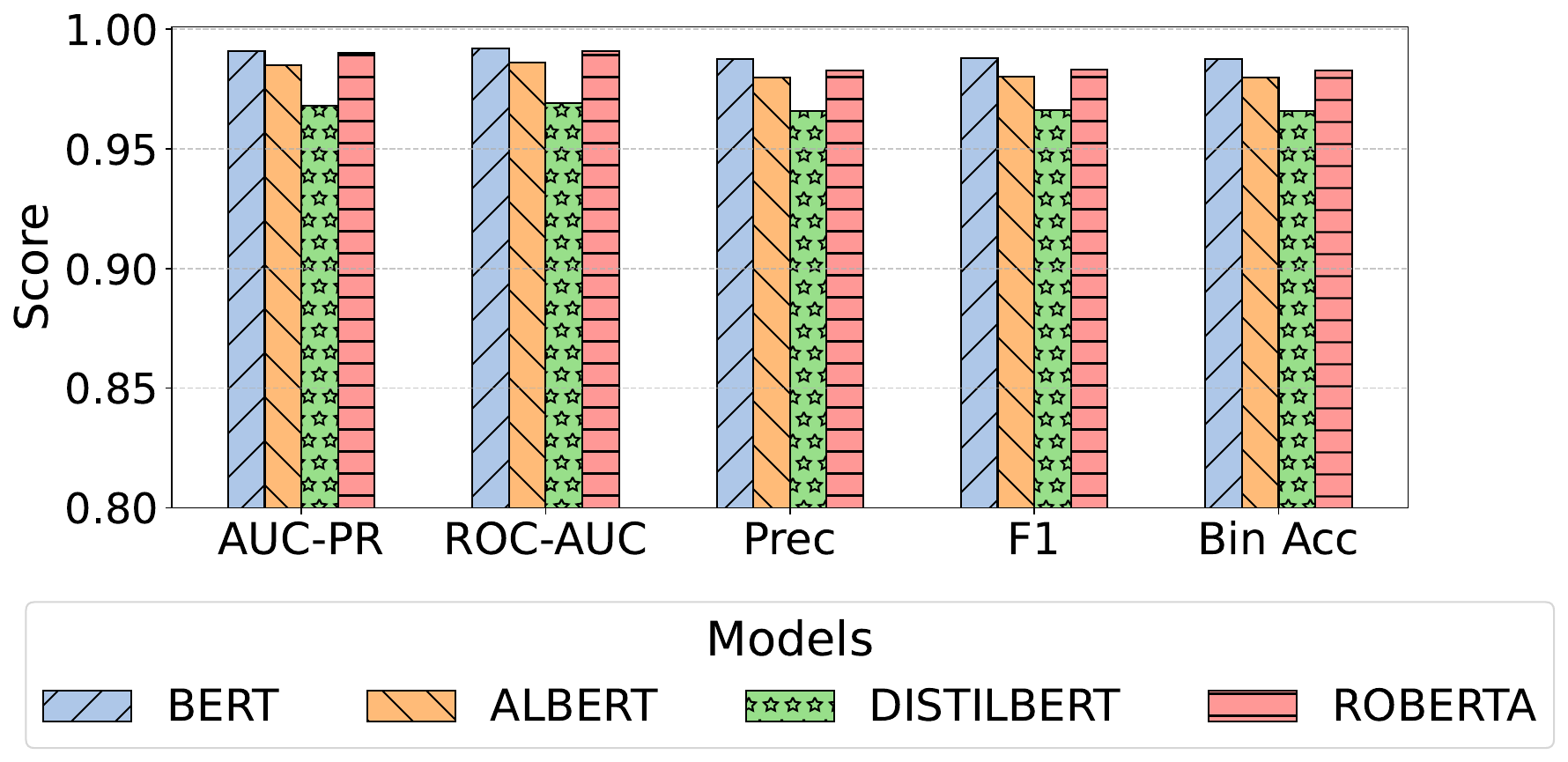}}
\hspace{0.03\textwidth}
\subfigure[{\scriptsize SSTI.}\label{fig:bert_ssti}]
{\includegraphics[width=0.48\textwidth]{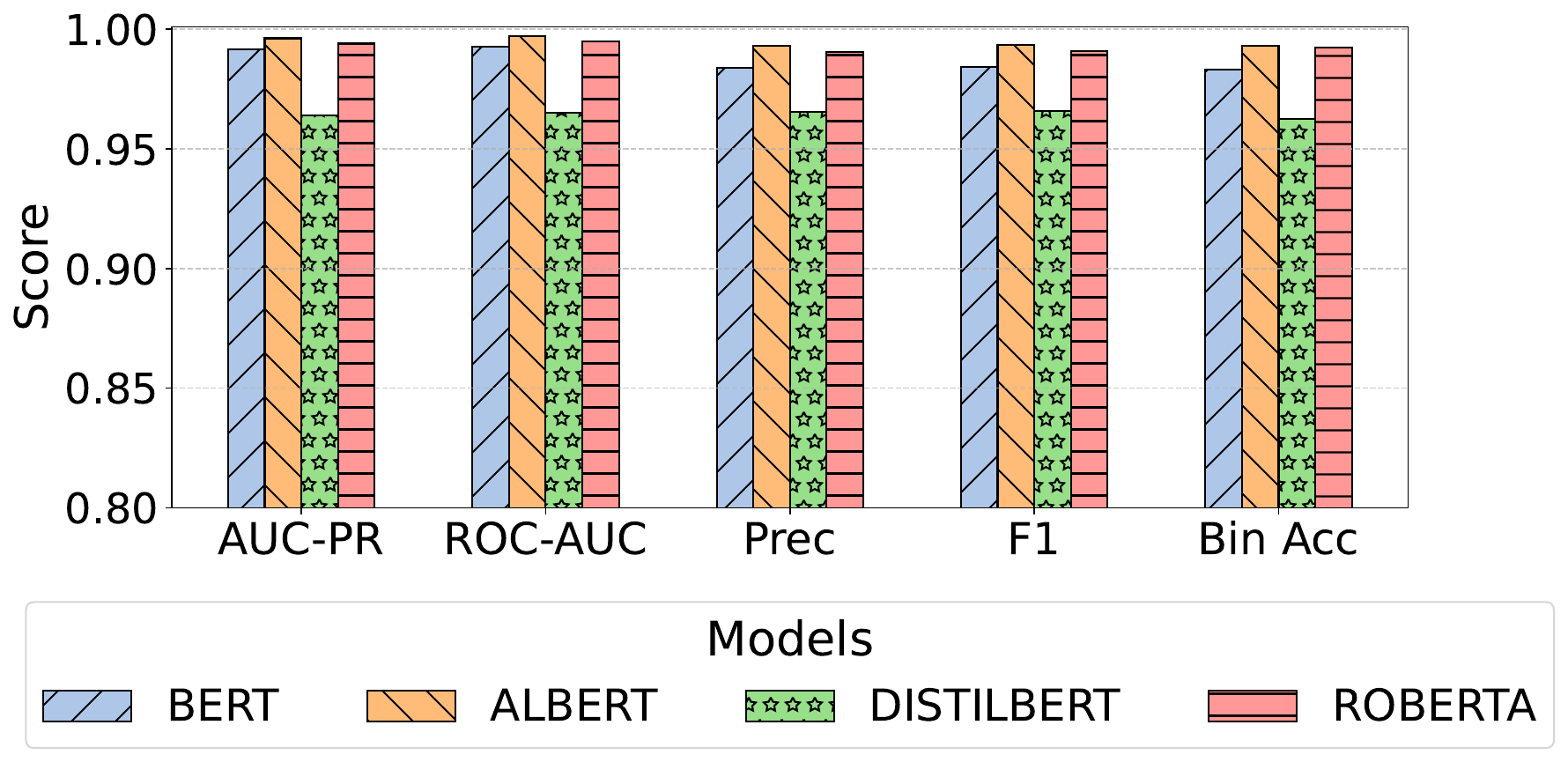}}
\hspace{0.03\textwidth}
\subfigure[{\scriptsize XSS.}\label{fig:bert_xss}]
{\includegraphics[width=0.48\textwidth]{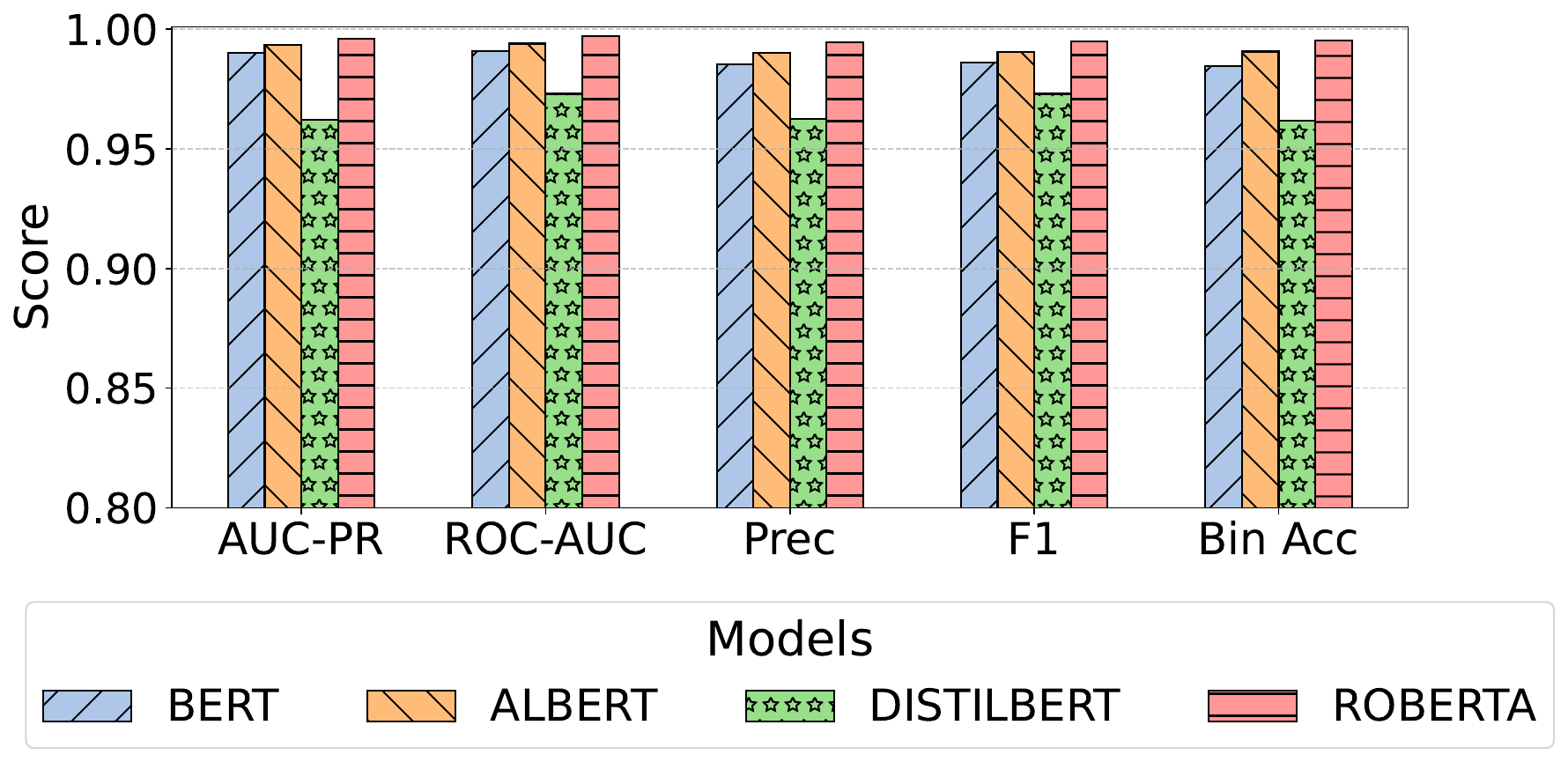}}

\vspace{0em}

\caption{Classification performance of BERT on different web vulnerabilities  using attack-specific training.}
\label{fig:bert}
\end{figure}

Figure~\ref{fig:bert} shows the classification performance of the BERT-based classifiers (\texttt{bert-base-uncased, albert-base-v2, distilbert-base-uncased, roberta-base}), used to build the classification tool.
All models  were trained and tested for each attack class considered.
Prior to model training, the datasets underwent preprocessing, which involved the removal of incomplete rows, balancing between positive and negative class. Subsequently, each dataset was partitioned into training and test subsets following an 80-20 random split.
All classifiers taken into account have been trained with identical hyperparameters: a maximum of 30 epochs and a batch size of 32. The optimizer is RectifiedAdam 
%(\texttt{tfa.optimizers.RectifiedAdam}) 
with a learning rate of $3 \times 10^{-5}$.
The loss function is binary cross-entropy.

The classifiers were evaluated by using 5 different metrics, three of which are based on the computation of the Area Under the Curve (AUC) for the
precision-recall (PR) and receiving operating characteristic (ROC). The other  three are the precision (Prec), the F1-score and the binary accuracy (Bin Acc).

During training, the best weights are saved using early stopping based on validation AUC through a model checkpoint callback.

As observed in Figure~\ref{fig:bert_sql}, the best performing model for the classification of  SQL Injection  is \texttt{bert-base-uncased}. \texttt{bert-base-uncased} demonstrates strong generalization on query patterns typical of this attack type. In the case of SSTI, shown in Figure~\ref{fig:bert_ssti}, the highest classification performance is achieved by \texttt{albert-base-v2}, indicating its greater sensitivity to the subtle templating syntax often involved in such injections. Finally, for XSS detection, depicted in Figure~\ref{fig:bert_xss}, the model that yields the best results is \texttt{roberta-base}, which appears to be particularly effective in capturing patterns associated with malicious HTML and JavaScript content.

These observations support the idea that different architectures may offer advantages for distinct types of attack vectors. %, further motivating the need for a specialized multi-classifier design.
This specialization strategy is motivated by principles of modular learning, which have shown improved generalization and interpretability in various domains~\cite{mixture_of_experts2017,modular_nlp2017}. 
By focusing each classifier on a single task, the system can better capture fine-grained patterns specific to each attack, reducing the risk of overgeneralization.

From a computational efficiency perspective, each classifier was trained on a NVIDIA A100 GPU with identical hyperparameters. 
On average across the three attack classes, the training time per model was approximately 2.1 hours, while dataset preparation 
(tokenization, balancing, and formatting) required less than 10 minutes per class. Once trained, inference proved highly efficient: 
The average prediction and report generation latency was 0.8 seconds per alert, ensuring the system operates efficiently and meets the time-sensitive demands of enterprise SOC environments.

% ****** Verificare******\textcolor{red}{aggingere per esempio ACC/f1 nel  migliore, forse dovremmo proporre una tabella riassuntiva Modello/attacco/performances . Dovremmo dire che dato l'alto numero di attacchi, anche qualche punto di probabilità in più puo fare la differenza. Mettiamo una tabella o un intervallo di confidenza}

\subsubsection{Model selection rationale and operational constraints}
Our evaluation focuses on open-weight encoder models (BERT, RoBERTa, ALBERT, DistilBERT) for two main reasons. 
First, the artifacts involved in our setting are predominantly textual (payloads, alerts, logs). Encoder architectures remain well suited for fine-grained, bidirectional pattern recognition needed to classify SQLi, XSS, and SSTI, while keeping the training process transparent and reproducible.

Second, open-weight encoders can be fine-tuned on curated security datasets. This capability is essential in a highly specialized and adversarial domain, where domain adaptation materially improves stability and accuracy. By contrast, proprietary frontier models (e.g., GPT-4) do not expose their weights and cannot be fine-tuned; they would need to be used in zero-shot or few-shot configurations only, which typically yields lower and less stable performance for cybersecurity tasks that require targeted adaptation. In addition, closed models entail data egress, cost, and governance constraints that are often incompatible with on-premise, regulated deployments.

% Third, the proposed contribution is architectural: an Agentic RAG framework that orchestrates specialized classifiers with iterative retrieval. The framework is model-agnostic and forward-compatible: more recent open-weight models (e.g., LLaMA 3, Mistral, Qwen2) can be integrated as drop-in replacements without structural changes to the system. This preserves the practical benefits of our current setup (low latency, on-prem deployment, reproducible training) while enabling future upgrades as models evolve.

% Looking ahead, we plan to unify the pipeline into a single decision flow with discrepancy analysis across classifier predictions, retrieved evidence, and raw telemetry. When inconsistencies arise, the agent will either iterate retrieval or escalate to human supervision, improving robustness while maintaining the latency and throughput required by enterprise SOC workflows.

\subsection{Context-Aware Orchestration via RAG}
\label{sec:results:rag_orch}
%While the use of specialized agents allows each classifier to focus on a specific attack type, ambiguity may still arise in edge cases. For example, multiple agents might assign similar probabilities to different attack types, or they might incorrectly detect an attack in a benign query due to pattern overlap. To handle such situations, we introduce a second layer of decision-making based on a RAG model acting as an orchestrator.

%Rather than selecting the label with the highest probability, which would be a naive winner-takes-all strategy, we leverage the RAG model to incorporate domain-specific knowledge and reason about the classification in context. This approach ensures that the final decision is not only based on numerical scores, but also informed by semantic evidence retrieved from a curated knowledge base.

The agentic-RAG receives a structured input consisting of the result of the query to the \textit{RAG} tool, the output probabilities from the three classifiers, and contextual instructions. 
It is tasked with verifying whether the most probable class is truly correct by checking for known signatures, keywords, and semantic patterns specific to each attack type.
%This general structure is instantiated dynamically depending on the class with the highest initial probability. 
In cases of conflicting scores or suspicious payloads that do not align well with any class, the RAG is capable of discarding false positives by evaluating the plausibility of the classification against known patterns.
We evaluated this architecture on a curated and unified subset of four popular attack datasets. 
Each query in the dataset was annotated with a dedicated attribute indicating the reference class label: \texttt{SSTI = 1}, \texttt{SQL Injection = 2}, and \texttt{XSS = 3}, and \texttt{0} to denote the absence of an attack.
These data were never used during the training phase of either the classifiers or the RAG model.

Let $f_{max}()$ be a function returning the index of the best classifier 
%starting from 1. If all classifiers are below $0.5$ the result of the function is $0$.
and %we define 
$c_i$ the probability returned from
%as the output of 
the $i^{th}$ classifier

\[
f_{\max}(c_1, \dots, c_N) =
\begin{cases}
\displaystyle \arg\max_{1 \leq i \leq N} \; \left\{c_i\right\}, & \text{if } \displaystyle \max_{1 \leq i \leq N} \left\{c_i\right\} \geq 0.5,\\[1ex]
0, & \text{otherwise.}
\end{cases}
\]

The initial evaluation using the classifier selected by $f_{max}(\dots)$ without RAG achieved an accuracy of $84.75\%$.
However, in some borderline cases, certain models incorrectly classified benign queries as attacks (false positives), likely due to structural similarities with malicious patterns. 
After integrating the RAG orchestrator, which reasons over context and retrieves relevant evidence, the classification accuracy increases significantly. The best accuracy performance ($94.92\%$) has been achieved using the \texttt{LLaMA3.1:8b} model and the \textit{RAG} tool. 
The following is an example of a prompt issued by the core LLM to the RAG tool.
%\vspace{0.8em}
\begin{tcolorbox}[before skip=10pt,  after skip=0pt, colback=gray!5, colframe=black!30, title=Generalized Prompt for RAG Decision]
\textbf{System:} This model analyzes suspicious queries and identifies the most likely web attack based on classifier outputs and contextual features.

\textbf{Inputs:} Query: \texttt{\{query\}}; Class probabilities: SQL Injection: \texttt{\{sql\_probability\}}, SSTI: \texttt{\{ssti\_probability\}}, XSS: \texttt{\{xss\_probability\}}.

\textbf{Task:} Analyze the query to identify patterns aligned with specific attacks (e.g., SQL keywords, HTML tags, template syntax). Determine the most semantically consistent class, justify the classification based on retrieved knowledge, and produce a class-specific feature vector.
\end{tcolorbox}
%\vspace{0.8em}

\subsection{Evaluation of Generated Explanations}
\label{sec:results:rag_eval}
To evaluate the quality of contextualized explanations generated by our RAG framework, we conducted an extensive benchmark across five different open-weight language models, all with approximately 7 billion parameters. The selected models are: \textbf{DeepSeek-R1 7B}~\cite{deepseekai2025}, \textbf{Gemma3:4B}~\cite{gemma2024}, \textbf{LLaMA3.1 8B}~\cite{touvron2024llama}, \textbf{Mistral 7B}~\cite{jiang2023mistral}, and \textbf{Qwen2.5 7B}~\cite{qwen2024}. All of these models were made available and executed within the \texttt{Ollama} environment~\cite{ollama2024}, enabling unified access and deployment for comparative evaluation\footnote{At the time the tests were conducted, these represented the state-of-the-art LLMs.}.

The framework is model-agnostic and forward-compatible: more recent open-weight models (e.g., LLaMA 4, Qwen3) can be integrated as drop-in replacements without structural changes to the system. % This preserves the practical benefits of our current setup (low latency, on-prem deployment, reproducible training) while enabling future upgrades as models evolve.

For each model, we generated explanations starting from attack payloads and their true class labels. Reference reports were manually curated, describing the attack context, the techniques employed, and key indicators embedded in the payloads. These served as the ground truth for our evaluation.

\subsubsection{Evaluation via Metrics}
The evaluation was designed to assess two core aspects: 
\begin{inparaenum}[\itshape(i)\upshape]
    \item the fidelity of the generated explanation with respect to the real nature of the attack and its classification, and 
    \item the semantic completeness and clarity of the descriptions.    
\end{inparaenum}

To this end, we employed a suite of well-established metrics from the natural language generation (NLG) literature. Specifically, we used \textbf{BLEU}~\cite{papineni2002bleu}, \textbf{ROUGE}~\cite{lin2004rouge}, and \textbf{METEOR}~\cite{banerjee2005meteor} for lexical overlap and surface-level comparison. For measuring semantic similarity, we adopted \textbf{BERTScore}~\cite{zhang2019bertscore}, which leverages contextual embeddings from pre-trained language models. Additionally, we introduced a custom factual consistency metric, designed to evaluate the alignment of generated explanations with relevant evidence retrieved by the system.

\begin{table*}[ht]
\centering
\caption{Automatic evaluation scores of RAG-generated explanations per model (average across all attack types).}
\label{tab:rag_eval_models}
\fontsize{8pt}{9pt}\normalfont{
\begin{tabular}{lccccc}
\toprule
\textbf{Model} & \textbf{BLEU} & \textbf{ROUGE} & \textbf{METEOR} & \textbf{BERTScore} & \textbf{Factual Consistency} \\
\midrule
DeepSeek-R1:7B     & 0.86 & 0.89 & 0.84 & 0.93 & 0.95 \\
Gemma3:4B          & 0.84 & 0.87 & 0.82 & 0.92 & 0.94 \\
LLaMA3.1:8B        & 0.88 & 0.90 & 0.86 & 0.94 & 0.96 \\
Mistral:7B         & 0.85 & 0.88 & 0.84 & 0.93 & 0.95 \\
Qwen2.5:7B         & 0.87 & 0.89 & 0.85 & 0.94 & 0.96 \\
\bottomrule
\end{tabular}
}
\end{table*}

Table~\ref{tab:rag_eval_models} reports the average scores per model, aggregated over explanations for three representative attack types: SQL Injection, Server-Side Template Injection (SSTI), and Cross-Site Scripting (XSS).
%
%As shown in Table~\ref{tab:rag_eval_models}, a
All models show strong performance in all evaluation metrics, with LLaMA3.1:8B and Qwen2.5:7B slightly outperforming the others in both semantic fidelity and factual alignment. %These findings confirm the viability of leveraging mid-sized open models within a RAG pipeline to generate reliable and interpretable threat descriptions.

\subsubsection{Evaluation via LLM-as-a-Judge}
\label{sec:results:llm_judge}
To complement automatic evaluation metrics, we employed a Large Language Model as an external ``judge'' to provide qualitative ratings of the generated explanations. Inspired by recent evaluation protocols~\cite{goyal2022news, liu2023gpteval}, this approach uses an independent model to simulate expert-level assessment of textual quality and domain relevance.

Importantly, to avoid bias due to internal feedback loops, we employed \textbf{GPT-4}~\cite{openai2023gpt4} as the external evaluator. This ensures the assessment is detached from the models used for generation and allows for a more objective comparison across outputs.

The evaluation consisted of two complementary components:
\begin{itemize}[(i)]
    \item \textbf{General Explanation Evaluation:} For each generated explanation, GPT-4 rated it on a 1-5 scale for \textit{clarity}, \textit{informativeness}, and \textit{semantic alignment} with the input payload and the attack label.
    \item \textbf{Attack-Specific Evaluation:} GPT-4 assessed each explanation based on how well it captured the characteristics of the corresponding attack type.

\end{itemize}

 The criteria were:
\begin{itemize}
    \item \textit{Pattern Recognition}: recognition of syntactic or structural cues (e.g., SQL keywords, template markers, script tags).
    \item \textit{Contextualization}: integration of the payload into a realistic scenario.
    \item \textit{Terminology Use}: precision and correctness of cybersecurity-related language.
\end{itemize}

The average scores for each model, aggregated across the three attack types, are shown in Table~\ref{tab:llm_judge_models}.

\begin{table*}[ht]
\centering
\caption{GPT-4 based LLM-judge scores (1-5 scale) per model.}
\label{tab:llm_judge_models}
\fontsize{8pt}{9pt}\normalfont{
\begin{tabular}{lcccc}
\toprule
\textbf{Model} & \textbf{Pattern Recognition} & \textbf{Contextualization} & \textbf{Terminology Use} & \textbf{Overall Avg.} \\
\midrule
DeepSeek-R1 7B     & 4.9 & 4.7 & 4.8 & 4.8 \\
Gemma3:4B          & 4.7 & 4.6 & 4.7 & 4.7 \\
LLaMA3.1:8B        & 5.0 & 4.8 & 4.9 & 4.9 \\
Mistral:7B         & 4.8 & 4.7 & 4.8 & 4.8 \\
Qwen2.5:7B         & 4.9 & 4.8 & 4.9 & 4.9 \\
\bottomrule
\end{tabular}
}
\end{table*}

The LLM-as-a-Judge results reinforce the trends observed in the automatic evaluations. LLaMA3.1:8B and Qwen2.5:7B received the highest scores in all dimensions, particularly for pattern recognition and precise technical terminology.

\subsection{Ablation Study: Impact of Retrieval}
\label{sec:results:ablation}
To quantify the contribution of the retrieval module to the overall system performance, we conducted an ablation study by comparing the quality of explanations produced with and without the use of RAG. Each explanation was rated on a scale from 1 to 5 by an LLM-judge based on criteria such as semantic completeness, accuracy, and grounding~\cite{goyal2022news, liu2023gpteval}.

\begin{figure}[!htb]
\centering
\includegraphics[width=0.45\textwidth]{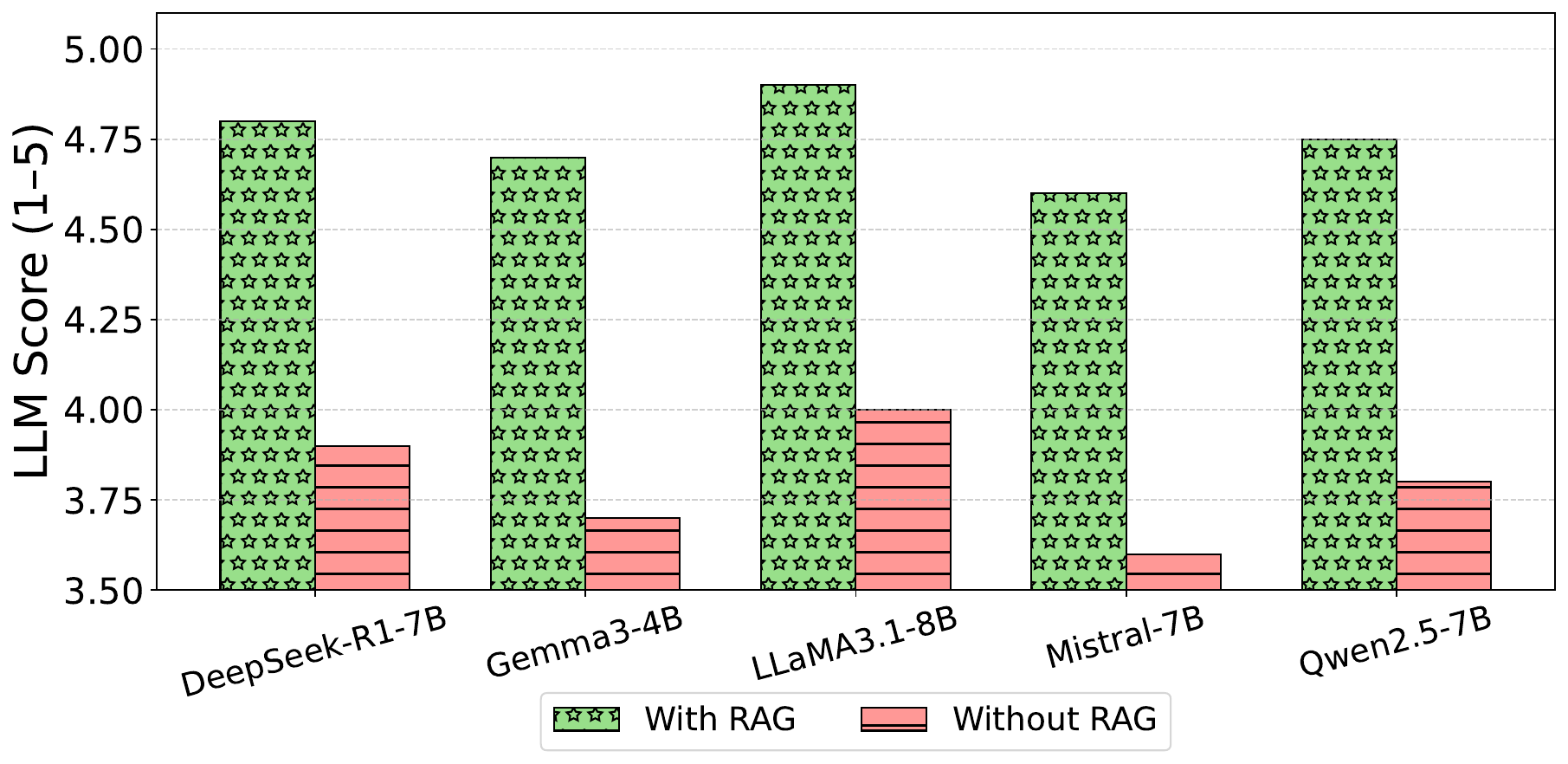}
\caption{Comparison between explanations generated with and without retrieval. LLM-based scoring shows consistent advantage from RAG-enhanced generation.}
\label{fig:ablation_study}
\end{figure}

As seen in Figure~\ref{fig:ablation_study}, explanations produced with RAG consistently score higher across all models, with \texttt{LLaMA3.1:8B} reaching the highest mean rating of 4.9. This confirms the added value of retrieval for producing grounded and high-quality justifications.

\subsection{Robustness Evaluation}
\label{sec:results:robustness}

To evaluate the robustness of the proposed RAG-based classification system, we designed a controlled experiment targeting two critical scenarios where machine learning models often fail:

\begin{inparaenum}[\itshape(i)\upshape]
    \item \textbf{Adversarial examples}, where input queries are subtly perturbed to simulate evasion attempts while preserving their original semantics (e.g., character obfuscation, encoding variations, or token reordering), and
    \item \textbf{Out-of-distribution (OOD) inputs}, consisting of queries drawn from attack categories that were not present during training (e.g., Path Traversal, Command Injection), thus requiring the model to reject or correctly classify novel patterns~\cite{ribeiro2020beyond,hendrycks2020pretrained}.
\end{inparaenum}

For each language model under evaluation (DeepSeek-R1:7B, Gemma3:4B, LLaMA3.1:8B, Mistral:7B, and Qwen2.5:7B), we constructed two benchmark sets:
\begin{itemize}
    \item A set of \textbf{100 adversarial examples} per attack category, crafted via controlled perturbation techniques.
    \item A set of \textbf{100 OOD queries}, selected from disjoint web attack datasets and annotated to indicate non-membership in the known classes.
\end{itemize}

The metric used is \textbf{Correct Classification (\%)}, which quantifies the model's ability to accurately identify or reject inputs under challenging conditions. For each model and scenario, the percentage is computed as:

\begin{equation}
\text{Correct Classification (\%)} = \frac{\text{\# Correct Predictions}}{\text{\# Predictions}} \times 100\%
\end{equation}

An adversarial input is considered correctly classified if the model assigns the true attack label despite the perturbation. An OOD query is considered correct if the model abstains from misclassifying it as one of the known attack types.

\begin{figure}[!htb]
\centering
\includegraphics[width=0.45\textwidth]{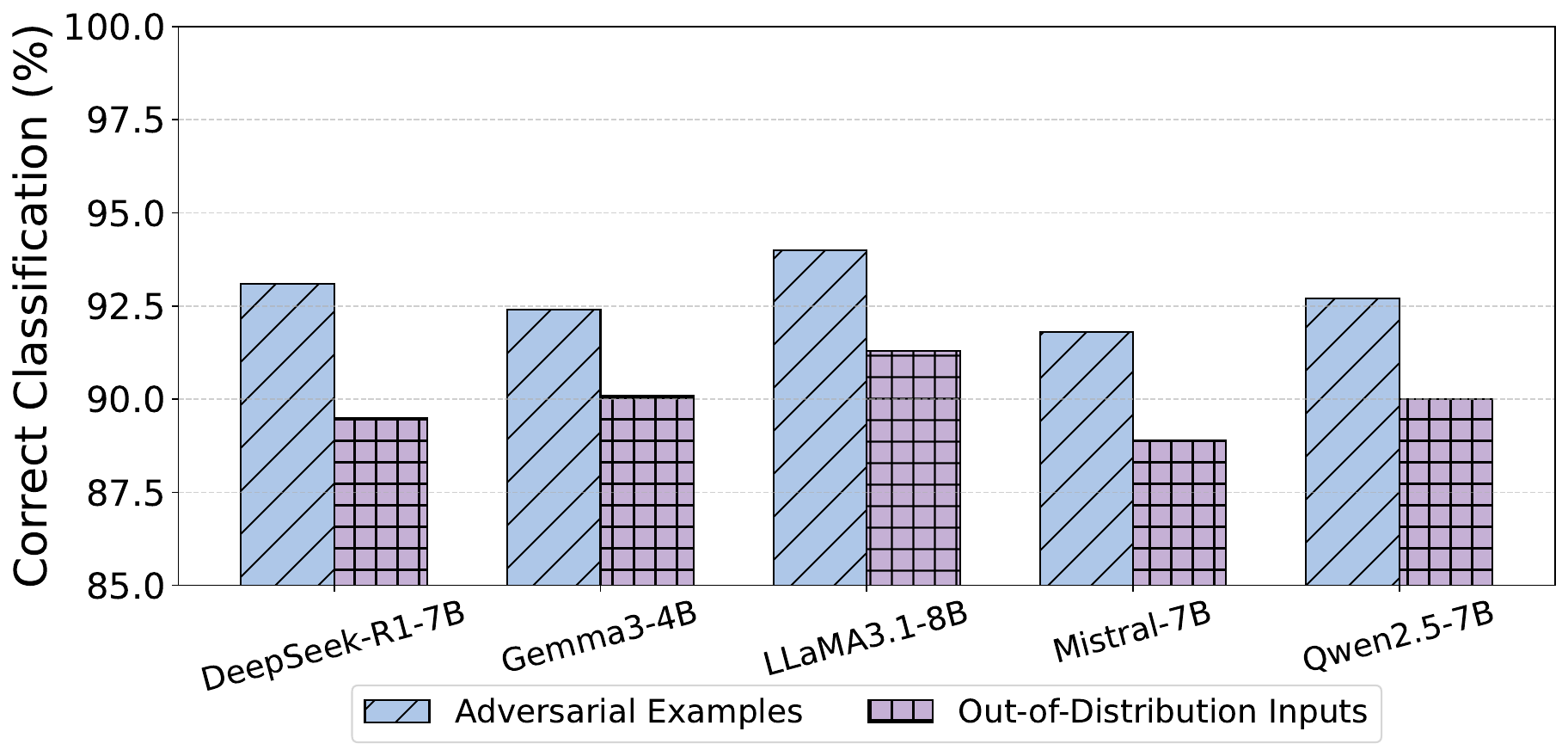}
\caption{Evaluation of model robustness based on the percentage of correct classifications under two conditions: (i) \textit{Adversarial Examples}, where inputs are perturbed to simulate evasion attacks, and (ii) \textit{Out-of-Distribution (OOD) Inputs}, representing unseen attack categories. The metric \textbf{Correct Classification (\%)} reflects the number of accurate predictions out of 100 test cases for each scenario.}
\label{fig:robustness_test}
\end{figure}

As shown in Figure~\ref{fig:robustness_test}, \texttt{LLaMA3.1:8B} achieves the highest robustness across both scenarios, correctly classifying 94\% of adversarial examples and 91\% of OOD inputs. \texttt{Mistral:7B} also shows strong performance under adversarial conditions (93\%), though with slightly reduced reliability on unseen categories (88.5\%). Other models such as \texttt{DeepSeek-R1:7B} and \texttt{Gemma3:4B} exhibit a larger performance gap between adversarial and OOD handling, suggesting room for improvement in generalization.

\vspace{1em}

\section{Discussion}
\label{sec:discussion}

% riprende e completa quanto detto in 5.1

As shown in Section~\ref{sec:results} the results obtained with \toolname{} demonstrate the effectiveness of a modular and Agentic RAG approach to the classification and explanation of cyber threats. Delegating the classification task to specialized models, each optimized for a specific attack type, proved more advantageous than a monolithic solution.

The usage of multiple classifiers allows for a modular architecture that contributes to the robustness of the system. If one classifier performs poorly due to data imbalance or ambiguity in one attack category, it does not compromise the performance of the others. This separation of responsibilities leads to a more reliable overall detection pipeline, reducing the likelihood of false positives and false negatives in critical scenarios.%~\cite{task_decomposition,multi_agent_security}.

Another advantage is the increased transparency of the classification process. Since each agent operates independently, it is possible to trace back which features or evidence were used to make a decision for each specific attack category. This supports explainability and makes the system more trustworthy in real-world deployments.

In contrast, training a single model to jointly classify all attack types resulted in significantly lower performance, with an overall accuracy of just $0.734$. This highlights the limitations of monolithic architectures in dealing with structurally diverse input distributions and further motivates the agent-based decomposition approach.

%riprendere 5.2
The general architecture of an Agentic RAG demonstrates the added value of combining probabilistic prediction with contextual and semantic understanding, ensuring a more accurate and explainable decision-making process in critical cybersecurity applications (Section~\ref{sec:results:rag_eval}).

This decoupled architecture enabled better discrimination capability and reduced misclassification, particularly in edge cases. The integration of the RAG component led to a substantial increase in accuracy, from 84.75\% to 94.92\%, highlighting its role as a semantic orchestrator.

A further significant contribution is the RAG component, which improves final decision-making by reasoning over retrieved evidence and domain knowledge. This mechanism not only improves classification accuracy, but also generates technically grounded and context-aware explanations. The synergy between a semantic search engine and a summarization-optimized LLM enables the production of comprehensive reports, as confirmed by both automatic metrics (BLEU, ROUGE, BERTScore) and GPT-4-based evaluation.

The use of a thematic knowledge base, organized by attack type and easily extensible with internal documentation (e.g., PDF files), ensures flexibility and adaptability. In addition, robustness tests under adversarial and out-of-distribution scenarios show resilience, which is critical for real-world deployments. The interactive chat interface further enhances usability by allowing analysts to query the system naturally, receive personalized feedback, and interact with the output according to their level of expertise.

One of the main limitations of this approach is its dependence on the quality of the data within the knowledge base and the precision of the classifier. A poorly maintained knowledge base or an inaccurate classifier can significantly impact the system's effectiveness.
In addition, the tool is not equipped to describe new types of attacks or attacks it hasn't been trained to recognize. This limitation makes it less effective against evolving cyber threats, as it can only provide adequate descriptions for attacks that its classifier has been specifically trained on.
The effectiveness of \toolname{} is directly dependent on the performance of the preceding IDS/IPS. If the IDS/IPS fails to recognize a security event as potentially dangerous, it won't generate a log entry for that event. Consequently, \toolname{} will have no data to analyze, rendering it useless in that specific scenario. This highlights a critical vulnerability in the system's pipeline, where a failure at the initial detection stage prevents any subsequent analysis.
We are aware of this limitation. However, given that \toolname{} is designed to operate in contexts with known threats and its effectiveness is dependent on the Knowledge Base, this issue can be mitigated.

It is sufficient to ensure that the rules for the IDS/IPS are constantly updated. To do this, dedicated online services can be used that offer regularly updated rules and signatures, so as to keep the intrusion prevention and detection system always up to date with the most recent and known threats.

Additional studies on detecting threats not based on a priori known signatures are already present in the literature, but they are considered outside the scope of this work.

% \textcolor{red}{Da rimpolpare o spostare}
Building on the approaches mentioned earlier, recent contributions have introduced specialized RAG-based frameworks tailored specifically for cybersecurity tasks. TechniqueRAG~\cite{lekssays-etal-2025-techniquerag} addresses the challenge of adversarial technique annotation in low-resource scenarios by combining off-the-shelf retrievers with instruction-tuned LLMs and a novel zero-shot re-ranking strategy. Its main strength lies in improving precision when labeled data are scarce, while \toolname{} focuses on real-time IDS/IPS alert triage and the generation of structured, SOC-ready reports. AURA~\cite{aura2025} introduces a multi-agent architecture for the attribution of APTs, integrating heterogeneous sources such as Trusted Third Parties (TTPs), Indicators of Compromise (IoCs), and malware artifacts. In contrast, \toolname{} targets a different stage of the defense pipeline, aiming to reduce false positives and provide interpretable explanations at the alert triage level. 
Another complementary direction is CTIKG~\cite{huang2024ctikg}, which enriches RAG pipelines with knowledge graph structures to enhance cyber threat intelligence and contextual reasoning. 
%Finally, frameworks such as LLM-ATTACKGUARD\cite{llm_attackguard2025} emphasize defensive robustness by filtering adversarial prompts and improving resilience of RAG-enabled SOC tools. 
%Collectively, these systems highlight the growing diversity of Agentic RAG methodologies in cybersecurity. \toolname{} differentiates itself by directly addressing operational needs in enterprise SOCs, with modular extensibility, specialized classifiers, and interpretable reporting that complements attribution and intelligence focused frameworks.

\section{Conclusion}
\label{sec:conclusion}

This work introduces a modular framework for cyber-attack classification and description by integrating a specialized \textit{Classification} tool with a context-aware \textit{RAG} component. \toolname{} merges the precision of fine-tuned large language models, trained to detect specific web-based malicious payload, with the abstraction capabilities and enriched contextual reasoning offered by the RAG pipeline.

Through extensive evaluations, covering both quantitative performance metrics and qualitative interpretability assessments, \toolname{} consistently outperforms traditional monolithic classifiers in terms of accuracy, robustness, and explainability. The framework is particularly effective in handling complex, noisy, or ambiguous inputs, and it demonstrates strong generalization capabilities across previously unseen payloads, highlighting its adaptability and reliability in dynamic cybersecurity contexts.

\toolname{} represents a step forward in intelligent automation for incident response, providing security analysts with a reliable, adaptable, and semantically enriched assistant capable of acting as a virtual cybersecurity expert. Its modular architecture supports seamless integration with existing security infrastructures, enabling organizations to enhance their detection and response workflows without major architectural changes. Furthermore, this modularity lays a solid foundation for future extensions, such as incorporating new analytical components, integrating with external threat intelligence sources, or adapting to evolving attack surfaces and organizational needs.

Looking ahead, future developments will aim to expand the scope of supported attack types by expanding the underlying taxonomy and refining detection capabilities. A key area of enhancement involves the integration of structured knowledge sources, such as knowledge graphs, to support more advanced, explainable, and context-aware reasoning. In parallel, efforts will be directed toward enabling controlled and auditable automated response mechanisms, allowing \toolname{} to act not only as an analytical assistant but also as an intelligent orchestrator of defensive actions. These advancements will encourage greater integration with Security Information and Event Management (SIEM) platforms and threat intelligence pipelines, ultimately leading to a more autonomous, scalable, and proactive cybersecurity defense architecture.
Overall, \toolname{} emerges as a robust, innovative, and forward-looking prototype, that sets the foundation for the next generation of intelligent cybersecurity tools, combining analytical rigor with actionable insight.


\section*{Acknowledgments}
This work was partially supported by the projects SERICS  (PE00000014) and FAIR (PE0000013) under the MUR National Recovery and Resilience Plan funded by the European Union - NextGenerationEU.

The work of Francesco A. Pironti was supported by Agenzia per la cybersicurezza nazionale under the 2024-2025 funding programme for promotion of XL cycle PhD research in cybersecurity (CUP H23C24000640005).

\section*{Declaration of generative AI and AI-assisted technologies in the writing process}
During the preparation of this work the authors used ChatGPT in order to check for grammar errors, typos, and overall writing clarity. After using this tool/service, the authors reviewed and edited the content as needed and take full responsibility for the content of the publication.

\balance
\bibliographystyle{elsarticle-num}
\bibliography{references}

\begin{thebibliography}{10}
\expandafter\ifx\csname url\endcsname\relax
  \def\url#1{\texttt{#1}}\fi
\expandafter\ifx\csname urlprefix\endcsname\relax\def\urlprefix{URL }\fi
\expandafter\ifx\csname href\endcsname\relax
  \def\href#1#2{#2} \def\path#1{#1}\fi

\bibitem{mohamed2025}
N.~Mohamed, Artificial intelligence and machine learning in cybersecurity: a deep dive into state-of-the-art techniques and future paradigms, Knowledge and Information Systems 67 (2025) 6969--7055.
\newblock \href {https://doi.org/10.1007/s10115-025-02429-y} {\path{doi:10.1007/s10115-025-02429-y}}.

\bibitem{LIAO2013}
H.-J. Liao, C.-H. {Richard Lin}, Y.-C. Lin, K.-Y. Tung, Intrusion detection system: A comprehensive review, Journal of Network and Computer Applications 36~(1) (2013) 16--24.
\newblock \href {https://doi.org/10.1016/j.jnca.2012.09.004} {\path{doi:10.1016/j.jnca.2012.09.004}}.

\bibitem{blefari2024}
F.~Blefari, F.~A. Pironti, A.~Furfaro, Toward a log-based anomaly detection system for cyber range platforms, in: The 19th International Conference on Availability, Reliability and Security (ARES 2024), ACM, 2024.
\newblock \href {https://doi.org/10.1145/3664476.3669976} {\path{doi:10.1145/3664476.3669976}}.

\bibitem{llm_cybersecurity_sota2024}
F.~N. Motlagh, M.~Hajizadeh, M.~Majd, P.~Najafi, F.~Cheng, C.~Meinel, Large language models in cybersecurity: State-of-the-art (2024).
\newblock \href {http://arxiv.org/abs/2402.00891} {\path{arXiv:2402.00891}}.

\bibitem{lewis2020retrieval}
P.~Lewis, E.~Perez, A.~Piktus, F.~Petroni, V.~Karpukhin, N.~Goyal, H.~K\"{u}ttler, M.~Lewis, W.-t. Yih, T.~Rockt\"{a}schel, S.~Riedel, D.~Kiela, Retrieval-augmented generation for knowledge-intensive nlp tasks, in: Proceedings of the 34th International Conference on Neural Information Processing Systems, NIPS '20, Curran Associates Inc., Red Hook, NY, USA, 2020.

\bibitem{yamin2024applications}
M.~Mudassar~Yamin, E.~Hashmi, M.~Ullah, B.~Katt, Applications of llms for generating cyber security exercise scenarios, IEEE Access 12 (2024) 143806--143822.
\newblock \href {https://doi.org/10.1109/ACCESS.2024.3468914} {\path{doi:10.1109/ACCESS.2024.3468914}}.

\bibitem{deng2025ai}
Z.~Deng, Y.~Guo, C.~Han, W.~Ma, J.~Xiong, S.~Wen, Y.~Xiang, Ai agents under threat: A survey of key security challenges and future pathways, ACM Comput. Surv. 57~(7) (Feb. 2025).
\newblock \href {https://doi.org/10.1145/3716628} {\path{doi:10.1145/3716628}}.

\bibitem{behera2025artificial}
A.~Behera, A.~Vedashree, M.~R. Kumar, K.~Upreti, Artificial intelligence and cybersecurity prospects and confronts, in: Navigating Cyber-Physical Systems With Cutting-Edge Technologies, IGI Global Scientific Publishing, 2025, pp. 155--184.
\newblock \href {https://doi.org/10.4018/979-8-3693-5728-6.ch006} {\path{doi:10.4018/979-8-3693-5728-6.ch006}}.

\bibitem{FURFARO2020}
A.~Furfaro, P.~Pace, A.~Parise, Facing {DDoS} bandwidth flooding attacks, Simulation Modelling Practice and Theory 98 (2020) 101984.
\newblock \href {https://doi.org/10.1016/j.simpat.2019.101984} {\path{doi:10.1016/j.simpat.2019.101984}}.

\bibitem{debar2009}
H.~Debar, An introduction to intrusion-detection systems (01 2009).

\bibitem{Angiulli2018}
F.~Angiulli, L.~Argento, A.~Furfaro, A.~Parise, A hierarchical hybrid framework for modelling anomalous behaviours, Simulation Modelling Practice and Theory 82 (2018) 103--115.
\newblock \href {https://doi.org/10.1016/j.simpat.2017.12.013} {\path{doi:10.1016/j.simpat.2017.12.013}}.

\bibitem{gupta2017}
S.~Gupta, B.~B. Gupta, Cross-{{Site Scripting}} ({{XSS}}) attacks and defense mechanisms: Classification and state-of-the-art, International Journal of System Assurance Engineering and Management 8~(1) (2017) 512--530.
\newblock \href {https://doi.org/10.1007/s13198-015-0376-0} {\path{doi:10.1007/s13198-015-0376-0}}.

\bibitem{HYDARA2015}
I.~Hydara, A.~B.~M. Sultan, H.~Zulzalil, N.~Admodisastro, Current state of research on cross-site scripting ({XSS}) – a systematic literature review, Information and Software Technology 58 (2015) 170--186.
\newblock \href {https://doi.org/10.1016/j.infsof.2014.07.010} {\path{doi:10.1016/j.infsof.2014.07.010}}.

\bibitem{Zhao2023}
Y.~Zhao, Y.~Zhang, M.~Yang, \href{https://www.usenix.org/conference/usenixsecurity23/presentation/zhao-yudi}{Remote code execution from {SSTI} in the sandbox: Automatically detecting and exploiting template escape bugs}, in: 32nd USENIX Security Symposium (USENIX Security 23), USENIX Association, Anaheim, CA, 2023, pp. 3691--3708.
\newline\urlprefix\url{https://www.usenix.org/conference/usenixsecurity23/presentation/zhao-yudi}

\bibitem{Singh2016}
J.~P. Singh, Analysis of sql injection detection techniques, Theoretical and Applied Informatics 28 (05 2016).
\newblock \href {https://doi.org/10.20904/281-2037} {\path{doi:10.20904/281-2037}}.

\bibitem{brown2020language}
T.~B. Brown, B.~Mann, N.~Ryder, M.~Subbiah, J.~Kaplan, P.~Dhariwal, A.~Neelakantan, P.~Shyam, G.~Sastry, A.~Askell, S.~Agarwal, A.~Herbert-Voss, G.~Krueger, T.~Henighan, R.~Child, A.~Ramesh, D.~M. Ziegler, J.~Wu, C.~Winter, C.~Hesse, M.~Chen, E.~Sigler, M.~Litwin, S.~Gray, B.~Chess, J.~Clark, C.~Berner, S.~McCandlish, A.~Radford, I.~Sutskever, D.~Amodei, Language models are few-shot learners, in: Proceedings of the 34th International Conference on Neural Information Processing Systems, NIPS '20, Curran Associates Inc., Red Hook, NY, USA, 2020.

\bibitem{devlin2019bert}
J.~Devlin, M.-W. Chang, K.~Lee, K.~Toutanova, {BERT}: Pre-training of deep bidirectional transformers for language understanding, in: J.~Burstein, C.~Doran, T.~Solorio (Eds.), Proc. of the 2019 Conference of the North {A}merican Chapter of the Association for Computational Linguistics: Human Language Technologies, Volume 1 (Long and Short Papers), Association for Computational Linguistics, Minneapolis, Minnesota, 2019, pp. 4171--4186.
\newblock \href {https://doi.org/10.18653/v1/N19-1423} {\path{doi:10.18653/v1/N19-1423}}.

\bibitem{radford2019language}
A.~Radford, J.~Wu, R.~Child, D.~Luan, D.~Amodei, I.~Sutskever, \href{https://cdn.openai.com/better-language-models/language_models_are_unsupervised_multitask_learners.pdf}{Language models are unsupervised multitask learners}, OpenAI Blog 1~(8) (2019) 9.
\newline\urlprefix\url{https://cdn.openai.com/better-language-models/language_models_are_unsupervised_multitask_learners.pdf}

\bibitem{vaswani2017attention}
A.~Vaswani, N.~Shazeer, N.~Parmar, J.~Uszkoreit, L.~Jones, A.~N. Gomez, L.~Kaiser, I.~Polosukhin, Attention is all you need, in: Proceedings of the 31st International Conference on Neural Information Processing Systems, NIPS'17, Curran Associates Inc., Red Hook, NY, USA, 2017, p. 6000–6010.

\bibitem{liu2019roberta}
Y.~Liu, M.~Ott, N.~Goyal, J.~Du, M.~Joshi, D.~Chen, O.~Levy, M.~Lewis, L.~Zettlemoyer, V.~Stoyanov, {RoBERTa}: A robustly optimized bert pretraining approach (2019).
\newblock \href {http://arxiv.org/abs/1907.11692} {\path{arXiv:1907.11692}}.

\bibitem{touvron2024llama}
H.~Touvron, T.~Lavril, G.~Izacard, X.~Martinet, M.-A. Lachaux, T.~Lacroix, B.~Rozière, N.~Goyal, E.~Hambro, F.~Azhar, A.~Rodriguez, A.~Joulin, E.~Grave, G.~Lample, Llama: Open and efficient foundation language models (2023).
\newblock \href {http://arxiv.org/abs/2302.13971} {\path{arXiv:2302.13971}}.

\bibitem{cosentino2024exploiting}
C.~Cosentino, M.~G{\"u}nd{\"u}z-C{\"u}re, F.~Marozzo, {\c{S}}.~{\"O}zt{\"u}rk-Birim, Exploiting large language models for enhanced review classification explanations through interpretable and multidimensional analysis, in: International Conference on Discovery Science, Springer, 2024, pp. 3--18.

\bibitem{cantini2024multi}
R.~Cantini, C.~Cosentino, F.~Marozzo, Multi-dimensional classification on social media data for detailed reporting with large language models, in: IFIP International Conference on Artificial Intelligence Applications and Innovations, Springer, 2024, pp. 100--114.

\bibitem{wooldridge:2009}
M.~Wooldridge, An Introduction to MultiAgent Systems, 2nd Edition, Wiley, 2009.

\bibitem{LLM_Agents_survey}
Z.~Xi, W.~Chen, X.~Guo, W.~He, Y.~Ding, B.~Hong, M.~Zhang, J.~Wang, S.~Jin, E.~Zhou, R.~Zheng, X.~Fan, X.~Wang, L.~Xiong, Y.~Zhou, W.~Wang, C.~Jiang, Y.~Zou, X.~Liu, Z.~Yin, S.~Dou, R.~Weng, W.~Qin, Y.~Zheng, X.~Qiu, X.~Huang, Q.~Zhang, T.~Gui, The rise and potential of large language model based agents: a survey, Science China Information Sciences 68 (2025).
\newblock \href {https://doi.org/10.1007/s11432-024-4222-0} {\path{doi:10.1007/s11432-024-4222-0}}.

\bibitem{LangChain}
H.~Chase, \href{https://github.com/langchain-ai/langchain}{Langchain} (October 2022).
\newline\urlprefix\url{https://github.com/langchain-ai/langchain}

\bibitem{LlamaIndex}
J.~Liu, \href{https://github.com/jerryjliu/llama\_index}{Llamaindex} (November 2022).
\newline\urlprefix\url{https://github.com/jerryjliu/llama\_index}

\bibitem{Langdroid}
P.~Chalasani, S.~Jha, \href{https://github.com/langroid/langroid}{Langdroid}.
\newline\urlprefix\url{https://github.com/langroid/langroid}

\bibitem{simmons2009avoidit}
C.~Simmons, C.~Ellis, S.~Shiva, D.~Dasgupta, Q.~Wu, {AVOIDIT}: A cyber attack taxonomy, Tech. rep., Office of Naval Research (ONR), supported under grant N00014-09-1-0752 (2009).

\bibitem{howard2007modern}
F.~Howard, Modern web attacks, Network Security 2008~(4) (2008) 13--15.
\newblock \href {https://doi.org/https://doi.org/10.1016/S1353-4858(08)70053-9} {\path{doi:https://doi.org/10.1016/S1353-4858(08)70053-9}}.

\bibitem{autoattacker2024}
J.~Xu, J.~W. Stokes, G.~McDonald, X.~Bai, D.~Marshall, S.~Wang, A.~Swaminathan, Z.~Li, Autoattacker: A large language model guided system to implement automatic cyber-attacks (2024).
\newblock \href {http://arxiv.org/abs/2403.01038} {\path{arXiv:2403.01038}}.

\bibitem{rag_cyberattack2024}
S.~Rajapaksha, R.~Rani, E.~Karafili, A {RAG-Based} question-answering solution for cyber-attack investigation and attribution, in: Computer Security. ESORICS 2024 International Workshops, Springer Nature Switzerland, Cham, 2025, pp. 238--256.

\bibitem{chatapt2024}
K.~Guru, \href{https://purl.stanford.edu/tk104mm3260}{Chatapt: Applying large language models (llms) for nation-state cyber attack attribution}, Stanford Digital Repository (2024).
\newline\urlprefix\url{https://purl.stanford.edu/tk104mm3260}

\bibitem{Anshuman2022}
A.~Singh, B.~B. Gupta, \href{https://doi.org/10.4018/IJSWIS.297143}{Distributed denial-of-service (ddos) attacks and defense mechanisms in various web-enabled computing platforms: Issues, challenges, and future research directions}, Int. J. Semant. Web Inf. Syst. 18~(1) (2022) 1–43.
\newblock \href {https://doi.org/10.4018/IJSWIS.297143} {\path{doi:10.4018/IJSWIS.297143}}.
\newline\urlprefix\url{https://doi.org/10.4018/IJSWIS.297143}

\bibitem{cashell2004economic}
B.~W. Cashell, W.~D. Jackson, M.~Jickling, B.~Webel, \href{https://digital.library.unt.edu/ark:/67531/metadc817913/}{The economic impact of cyber-attacks}, Tech. rep., CRS Report for Congress (2004).
\newline\urlprefix\url{https://digital.library.unt.edu/ark:/67531/metadc817913/}

\bibitem{jovicic2006common}
B.~Jovicic, D.~Simic, Common web application attack types and security using {ASP.NET}, Comput. Sci. Inf. Syst. 3~(2) (2006) 83--96.
\newblock \href {https://doi.org/10.2298/CSIS0602083J} {\path{doi:10.2298/CSIS0602083J}}.

\bibitem{generative_ai_cybersecurity2024}
M.~A. Ferrag, F.~Alwahedi, A.~Battah, B.~Cherif, A.~Mechri, N.~Tihanyi, T.~Bisztray, M.~Debbah, Generative ai in cybersecurity: A comprehensive review of llm applications and vulnerabilities, Internet of Things and Cyber-Physical Systems 5 (2025) 1--46.
\newblock \href {https://doi.org/10.1016/j.iotcps.2025.01.001} {\path{doi:10.1016/j.iotcps.2025.01.001}}.

\bibitem{llm_cybersecurity_review2024}
J.~Zhang, H.~Bu, H.~Wen, Y.~Liu, H.~Fei, R.~Xi, L.~Li, Y.~Yang, H.~Zhu, D.~Meng, When {LLMs} meet cybersecurity: a systematic literature review, Cybersecurity 8~(1) (2025) 55.
\newblock \href {https://doi.org/10.1186/s42400-025-00361-w} {\path{doi:10.1186/s42400-025-00361-w}}.

\bibitem{CHEN2024104016}
Y.~Chen, M.~Cui, D.~Wang, Y.~Cao, P.~Yang, B.~Jiang, Z.~Lu, B.~Liu, A survey of large language models for cyber threat detection, Computers \& Security 145 (2024) 104016.
\newblock \href {https://doi.org/https://doi.org/10.1016/j.cose.2024.104016} {\path{doi:https://doi.org/10.1016/j.cose.2024.104016}}.

\bibitem{carbonell1998use}
J.~Carbonell, J.~Goldstein, The use of {MMR}, diversity-based reranking for reordering documents and producing summaries, in: Proceedings of the 21st annual international ACM SIGIR conference on Research and development in information retrieval, 1998, pp. 335--336.
\newblock \href {https://doi.org/10.1145/290941.291025} {\path{doi:10.1145/290941.291025}}.

\bibitem{Reimers2019SentenceBERTSE}
N.~Reimers, I.~Gurevych, Sentence-{BERT}: Sentence embeddings using {S}iamese {BERT}-networks, in: K.~Inui, J.~Jiang, V.~Ng, X.~Wan (Eds.), Proceedings of the 2019 Conference on Empirical Methods in Natural Language Processing and the 9th International Joint Conference on Natural Language Processing (EMNLP-IJCNLP), Association for Computational Linguistics, Hong Kong, China, 2019, pp. 3982--3992.
\newblock \href {https://doi.org/10.18653/v1/D19-1410} {\path{doi:10.18653/v1/D19-1410}}.

\bibitem{douze2025faisslibrary}
M.~Douze, A.~Guzhva, C.~Deng, J.~Johnson, G.~Szilvasy, P.-E. Mazaré, M.~Lomeli, L.~Hosseini, H.~Jégou, The faiss library (2025).
\newblock \href {http://arxiv.org/abs/2401.08281} {\path{arXiv:2401.08281}}.

\bibitem{XSS}
Kaggle, Xss dataset, \url{https://www.kaggle.com/datasets/syedsaqlainhussain/cross-site-scripting-xss-dataset-for-deep-learning} (2025).

\bibitem{sstidataset}
F.~Blefari, F.~A. Pironti, {SSTI} dataset, \url{https://github.com/francescopirox/ssti_dataset} (2025).

\bibitem{sqlinj}
Kaggle, Sql injection dataset, \url{https://www.kaggle.com/datasets/syedsaqlainhussain/sql-injection-dataset/data} (2025).

\bibitem{mixture_of_experts2017}
N.~Shazeer, A.~Mirhoseini, K.~Maziarz, A.~Davis, Q.~Le, G.~Hinton, J.~Dean, Outrageously large neural networks: The sparsely-gated mixture-of-experts layer (2017).
\newblock \href {http://arxiv.org/abs/1701.06538} {\path{arXiv:1701.06538}}.

\bibitem{modular_nlp2017}
J.~Andreas, D.~Klein, S.~Levine, Modular multitask reinforcement learning with policy sketches, in: Proceedings of the 34th International Conference on Machine Learning - Volume 70, ICML'17, JMLR.org, 2017, p. 166–175.

\bibitem{deepseekai2025}
DeepSeek-AI, Deepseek-r1: Incentivizing reasoning capability in llms via reinforcement learning (2025).
\newblock \href {http://arxiv.org/abs/2501.12948} {\path{arXiv:2501.12948}}.

\bibitem{gemma2024}
{Google DeepMind}, \href{https://ai.google.dev/gemma}{Gemma: Open models based on gemini research and technology}, accessed: 2024-03 (2024).
\newline\urlprefix\url{https://ai.google.dev/gemma}

\bibitem{jiang2023mistral}
A.~Q. Jiang, A.~Sablayrolles, A.~Mensch, C.~Bamford, D.~S. Chaplot, D.~de~las Casas, F.~Bressand, G.~Lengyel, G.~Lample, L.~Saulnier, L.~R. Lavaud, M.-A. Lachaux, P.~Stock, T.~L. Scao, T.~Lavril, T.~Wang, T.~Lacroix, W.~E. Sayed, Mistral 7b (2023).
\newblock \href {http://arxiv.org/abs/2310.06825} {\path{arXiv:2310.06825}}.

\bibitem{qwen2024}
A.~Cloud, \href{https://huggingface.co/Qwen/Qwen2-7B}{Qwen2: A family of open-source language models by alibaba cloud} (2024).
\newline\urlprefix\url{https://huggingface.co/Qwen/Qwen2-7B}

\bibitem{ollama2024}
{Ollama Project}, \href{https://ollama.com}{Ollama: Run llms locally} (2024).
\newline\urlprefix\url{https://ollama.com}

\bibitem{papineni2002bleu}
K.~Papineni, S.~Roukos, T.~Ward, W.-J. Zhu, Bleu: a method for automatic evaluation of machine translation, in: Proceedings of the 40th Annual Meeting on Association for Computational Linguistics, ACL '02, Association for Computational Linguistics, USA, 2002, p. 311–318.
\newblock \href {https://doi.org/10.3115/1073083.1073135} {\path{doi:10.3115/1073083.1073135}}.

\bibitem{lin2004rouge}
C.-Y. Lin, \href{https://aclanthology.org/W04-1013/}{Rouge: A package for automatic evaluation of summaries}, in: Proc. of Workshop on Text Summarization Branches Out, 2004, pp. 74--81.
\newline\urlprefix\url{https://aclanthology.org/W04-1013/}

\bibitem{banerjee2005meteor}
A.~Lavie, A.~Agarwal, \href{https://aclanthology.org/W07-0734/}{Meteor: an automatic metric for mt evaluation with high levels of correlation with human judgments}, in: Proc. of the Second Workshop on Statistical Machine Translation, StatMT '07, Association for Computational Linguistics, USA, 2007, p. 228–231.
\newline\urlprefix\url{https://aclanthology.org/W07-0734/}

\bibitem{zhang2019bertscore}
T.~Zhang, V.~Kishore, F.~Wu, K.~Q. Weinberger, Y.~Artzi, Bertscore: Evaluating text generation with bert, in: International Conference on Learning Representations (ICLR), 2020.

\bibitem{goyal2022news}
T.~Goyal, J.~J. Li, G.~Durrett, News summarization and evaluation in the era of {GPT-3} (2023).
\newblock \href {http://arxiv.org/abs/2209.12356} {\path{arXiv:2209.12356}}.

\bibitem{liu2023gpteval}
Y.~Liu, D.~Iter, Y.~Xu, S.~Wang, R.~Xu, C.~Zhu, {G}-eval: {NLG} evaluation using gpt-4 with better human alignment, in: H.~Bouamor, J.~Pino, K.~Bali (Eds.), Proc. of the 2023 Conference on Empirical Methods in Natural Language Processing, Association for Computational Linguistics, Singapore, 2023, pp. 2511--2522.
\newblock \href {https://doi.org/10.18653/v1/2023.emnlp-main.153} {\path{doi:10.18653/v1/2023.emnlp-main.153}}.

\bibitem{openai2023gpt4}
OpenAI, Gpt-4 technical report, \url{https://openai.com/research/gpt-4} (2023).

\bibitem{ribeiro2020beyond}
M.~T. Ribeiro, T.~Wu, C.~Guestrin, S.~Singh, Beyond accuracy: Behavioral testing of {NLP} models with {C}heck{L}ist (2020) 4902--4912\href {https://doi.org/10.18653/v1/2020.acl-main.442} {\path{doi:10.18653/v1/2020.acl-main.442}}.

\bibitem{hendrycks2020pretrained}
D.~Hendrycks, X.~Liu, E.~Wallace, A.~Dziedzic, R.~Krishnan, D.~Song, Pretrained transformers improve out-of-distribution robustness, in: D.~Jurafsky, J.~Chai, N.~Schluter, J.~Tetreault (Eds.), Proceedings of the 58th Annual Meeting of the Association for Computational Linguistics, Association for Computational Linguistics, Online, 2020, pp. 2744--2751.
\newblock \href {https://doi.org/10.18653/v1/2020.acl-main.244} {\path{doi:10.18653/v1/2020.acl-main.244}}.

\bibitem{lekssays-etal-2025-techniquerag}
A.~Lekssays, U.~Shukla, H.~T. Sencar, M.~R. Parvez, {T}echnique{RAG}: Retrieval augmented generation for adversarial technique annotation in cyber threat intelligence text, in: W.~Che, J.~Nabende, E.~Shutova, M.~T. Pilehvar (Eds.), Findings of the Association for Computational Linguistics: ACL 2025, Association for Computational Linguistics, Vienna, Austria, 2025, pp. 20913--20926.
\newblock \href {https://doi.org/10.18653/v1/2025.findings-acl.1076} {\path{doi:10.18653/v1/2025.findings-acl.1076}}.

\bibitem{aura2025}
N.~Rani, S.~K. Shukla, {AURA}: A multi-agent intelligence framework for knowledge-enhanced cyber threat attribution (2025).
\newblock \href {http://arxiv.org/abs/2506.10175} {\path{arXiv:2506.10175}}.

\bibitem{huang2024ctikg}
L.~Huang, X.~Xiao, \href{https://openreview.net/forum?id=DOMP5AgwQz}{{CTIKG}: {LLM}-powered knowledge graph construction from cyber threat intelligence}, in: First Conference on Language Modeling, 2024.
\newline\urlprefix\url{https://openreview.net/forum?id=DOMP5AgwQz}

\end{thebibliography}

\end{document}